\documentclass[a4paper,11pt]{article}
\pdfoutput=1 % if your are submitting a pdflatex (i.e. if you have
\usepackage{jcappub} % for details on the use of the package, please
\usepackage[T1]{fontenc} % if needed
\usepackage[utf8]{inputenc}
\usepackage{graphicx}
\usepackage{txfonts}
\usepackage[usenames,dvipsnames]{xcolor}
\usepackage{tabularx}
\newcolumntype{L}[1]{>{\raggedright\arraybackslash}p{#1}}
\newcolumntype{C}[1]{>{\centering\arraybackslash}p{#1}}
\newcolumntype{R}[1]{>{\raggedleft\arraybackslash}p{#1}}
\usepackage{multicol}
\usepackage{multirow}
\usepackage{makecell}
\usepackage{cancel}

\usepackage{amstext}
\usepackage{subfig}
\usepackage{enumitem}

\usepackage[integrals]{wasysym}

\usepackage{mathtools}

\usepackage[normalem]{ulem} %eduardo

\newcommand{\hi}{{H{\sc i}} }

\usepackage{caption}
\title{\boldmath Detectability of the 21\,cm signal with BINGO through cross-correlation with photometric surveys}

\author[a]{Gabriel A. S. Silva\note{Corresponding author.},}
\author[a]{Camila P. Novaes,}
\author[a]{Carlos A. Wuensche,}
\author[a]{Eduardo J. de Mericia,}
\author[a,b]{Bruno B. Bizarria,}
\author[c,d]{Jiajun Zhang,}
\author[e,f,g]{Elcio Abdalla,}
\author[e,h,i,j]{Filipe B. Abdalla,}
\author[k]{Amilcar R. Queiroz,}
\author[a,l,m]{Thyrso Villela,}
\author[n,o]{Bin Wang,}
\author[h,i,p]{Chang Feng,}
\author[q]{Edmar C. Gurjão,}
\author[h]{Alessandro Marins}

\affiliation[a]{Instituto Nacional de Pesquisas Espaciais, \\ Av. dos Astronautas 175, Jardim da Granja, São José dos Campos, SP, Brazil}
\affiliation[b]{Jodrell Bank Centre for Astrophysics, University of Manchester, \\ Manchester M13 9PL, UK}
\affiliation[c]{Shanghai Astronomical Observatory (SHAO), Nandan Road 80, Shanghai, 200030, China}
\affiliation[d]{State Key Laboratory of Radio Astronomy and Technology, 20A Datun Road, Chaoyang District, Beĳing, 100101, China}
\affiliation[e]{Instituto de Física, Universidade de São Paulo,\\ C.P. 66.318, CEP 05315-970, São Paulo, Brazil}
\affiliation[f]{Universidade Estadual da Paraíba,\\ Rua Baraúnas, 351, Bairro Universitário, Campina Grande, Brazil}
\affiliation[g]{Departamento de Física, Centro de Ciências Exatas e da Natureza, Universidade Federal da Paraíba,\\ CEP 58059-970, João Pessoa, Brazil}
\affiliation[h]{Department of Astronomy, School of Physical Sciences, University of Science and Technology of China,\\  Hefei, Anhui 230026, China}
\affiliation[i]{School of Astronomy and Space Science, University of Science and Technology of China,\\ Hefei,  Anhui 230026, China}
\affiliation[j]{Department of Physics and Electronics, Rhodes University,\\ PO Box 94, Grahamstown, 6140, South Africa}
\affiliation[k]{Unidade Acadêmica de Física, Univ. Federal de Campina Grande,\\ R. Aprígio Veloso, 58429-900 - Campina Grande, Brazil}
\affiliation[l] {Centro de Gest\~ao e Estudos Estrat\'egicos SCS Qd 9, Lote C, Torre C S/N Salas 401 a 405, 70308-200 - Bras\'ilia, DF, Brazil}
\affiliation[m] {Instituto de F\'{i}sica, Universidade de Bras\'{i}lia, Campus Universit\'ario Darcy Ribeiro, 70910-900 - Bras\'{i}lia, DF, Brazil }
\affiliation[n]{Center for Gravitation and Cosmology, Yangzhou University,\\ Yangzhou 224009, China}
\affiliation[o]{School of Aeronautics and Astronautics, Shanghai Jiao Tong University,\\ Shanghai 200240, China}
\affiliation[p]{CAS Key Laboratory for Research in Galaxies and Cosmology, University of Science and Technology of China,\\ Hefei, Anhui 230026, China}
\affiliation[q]{Departamento de Engenharia Elétrica, Univ. Federal de Campina Grande,\\ R. Aprígio Veloso, 58429-900 - Campina Grande,
Brazil}

\emailAdd{gabriel.souza@inpe.br}

\abstract{
21\,cm intensity mapping (\hi IM) can efficiently survey large cosmic volumes with good redshift resolution, but systematics and foreground contamination pose major challenges for extracting precise cosmological information. Cross-correlation with galaxy surveys offers an efficient mitigation strategy, as the two datasets have largely uncorrelated systematics.
We assess the detectability of the 21\,cm signal from the BINGO radio telescope through cross-correlation with the LSST photometric survey, given their strong overlap in area and redshift. Using lognormal simulations, we model the cosmological signal in the BINGO frequency range (980–1260\,MHz), incorporating thermal noise, foregrounds, and cleaning. LSST simulations include photometric redshift (photo-$z$) uncertainties and galaxy number density in the first three redshift bins ($\bar{z} \approx 0.25$, $0.35$, $0.45$), matching expected performance after $10$ years of the survey. 
We show that photo-$z$ errors significantly increase noise in the cross-correlation, reducing its statistical significance to levels comparable with the autocorrelation. Still, the \hi signal remains detectable through cross-correlation, even with LSST-like photo-$z$ uncertainties. Our results support the viability of this approach under realistic conditions and motivate further refinement of current analysis methods.

\qquad


}

\begin{document}

\maketitle
\flushbottom

\section{Introduction}
\label{SecIntroduction} 

The mapping of the distribution of the luminous matter has significantly contributed to our knowledge of the contents and evolution of the Universe's large-scale structure, while corroborating the validity of the standard $\Lambda$CDM model. 
However, even with excellent agreement with many observations, the $\Lambda$CDM model still fails in explaining most of the Universe's content, the so-called dark matter and dark energy. 
Motivated by the necessity of understanding the physical nature of these components, as well as addressing and explaining many other open questions in cosmology, a next generation of experiments is expected to perform cosmological measurements with unprecedented accuracy. 
Examples of these experiments are the James Webb Space Telescope (e.g., \citep{2022:JamesWebb-Robertson}), the Euclid satellite (e.g., \citep{amendola2018cosmology}), and the Vera C. Rubin Observatory’s Legacy Survey of Space and Time - LSST (e.g., \citep{ivezic2019lsst}), which surveys the galaxy distribution of the Universe through photometric and spectroscopic measurements in the optical and near-infrared spectra. 

As an alternative, we have also been using radio wavelengths to measure the clustering of galaxies through the redshifted 21\,cm emission resulting from the hyperfine transition line of neutral hydrogen (H{\sc i}). 
In particular, the intensity mapping (IM) of the 21\,cm signal has emerged as a promising way to map the three-dimensional distribution of structures in the Universe. 
Among the forthcoming 21\,cm intensity mapping instruments, we can mention Tianlai\footnote{\url{http://tianlai.bao.ac.cn/wiki/index.php/Main_Page}} \citep{chen2012tianlai}, HIRAX \citep{2022/hirax}, the Five-Hundred-Meter Aperture Spherical Radio Telescope \cite[FAST;][]{2011/fast}, the Square Kilometer Array Observatory\footnote{\url{https://www.skatelescope.org/}} \citep[SKAO;][]{2020/ska}, and  the Baryon Acoustic Oscillations from Integrated Neutral Gas Observations\footnote{\url{https://www.bingotelescope.org/en/}} \citep[BINGO;][]{battye2013h, 2015/bigot-sazy, abdalla2022bingoI}. While observations are already being conducted by, e.g., MeerKAT\footnote{\url{https://www.ska.ac.za/science-engineering/meerkat}} \citep{2017/santos}, a pathfinder survey and precursor of SKAO, and the Canadian Hydrogen Intensity Mapping Experiment\footnote{\url{https://chime-experiment.ca/}} \citep[CHIME;][]{2014/chime}.

Measuring the 21\,cm signal from galaxy clustering is a challenging task. The presence of strong galactic emission, with synchrotron radiation being the most intense component, along with the existence of extragalactic sources (both unresolved and resolved), act as foregrounds for 21\,cm observations \citep{battye2013h}. 
Separating these different components and extracting the desired cosmological signal is a crucial step, with many methods developed to recover the 21\,cm signal \citep{Tegmark:2003, alonso2015blind, olivari2016extracting, 2020/carucci}. 
Nevertheless, even when well reconstructed, the foreground-cleaned maps still contain residual contamination and a loss of the cosmological signal. 
Furthermore, systematic effects are also important sources of contamination, including thermal noise \citep{battye2013h, wuensche2022bingo}, $1/f$ noise \citep{2018/harper, 2021/li}, radio frequency interference, beam modeling \citep{ding_beam:2024}, and calibration errors \citep{Wang_calibration:2021}.
A way of mitigating such effects on power spectrum estimations and increasing the signal-to-noise ratio (SNR) is to estimate the cross-correlation signal between the foreground cleaned maps and galaxy catalogs, as the systematics and contaminant signal from the two datasets are uncorrelated. 
In fact, most of the power spectrum detection so far have been achieved via cross-correlation \citep{Chang:2010, switzer2013determination, masui2013measurement, Wolz_gbt:2022, cunnington2023h, amiri2023detection}, with a claim of detection of the autocorrelation from interferometric observations by \cite{paul2023first}.

Current correlated clustering studies, as reported above, rely on the use of power spectrum estimators in the Fourier space, given that the small surveyed areas (typically less than a few hundred deg$^2$ \citep{dodelson2003modern, gao2023asymptotic, carucci2024hydrogen}) %e.g., the $\approx 200$ deg$^2$ used by \cite{cunnington2023h} and $\approx 4$ deg$^2$ by \cite{mazumder2025hi}) 
allow the resampling of pixelized maps onto a Cartesian grid without accounting for wide-angle effects like non-parallelism on the line-of-sights for $\ell \gtrsim 100$ \citep{Blake:2019, cunnington2024accurate, benabou2024wide}. 
Recently, \cite{amiri2023detection} reported a detection of the 21\,cm signal using a stacking approach, averaging the 21\,cm maps at the locations of the objects in a galaxy catalog. 
To our knowledge, no detection was reported so far in the harmonic space, which has been less explored in this context. 

The analysis using the angular power spectrum has been employed primarily for Fisher matrix analyses, as in  \cite{Pourtsidou:2015, 2019/padmanabhan, padmanabhan2020cross}, where the authors forecasts astrophysical and cosmological parameters using auto- and cross-correlations of 21\,cm and other surveys. 
Less commonly, \citep{shi2020hir4} employ synthetic data to study the correlation between 21\,cm map and a galaxy survey.  
Regardless of the estimator, detections of the \hi signal through cross-correlation analyses have been performed using spectroscopic galaxy surveys, owing to their more precise redshift determination. 
In fact, most studies in the literature, including those based on simulations, follow this approach. An exception is presented by \cite{cunnington2019impact}, which investigates how the redshift uncertainty impacts the cross-correlation signal when using photometric surveys.

Previous work has highlighted the limitations of cross-correlation between \hi IM and photometric galaxy catalogs, such as difficulties in modeling the accurate selection function and the effect of smearing the signal in the line-of-sight direction, which produces a loss in the amplitude of the correlation signal \citep{cunnington2019impact}. 
However, due to the great possibilities that arise from current and future photometric galaxy surveys like the Dark Energy Survey (DES) \citep{2019/abbott-des}, LSST / Rubin Observatory \citep{ivezic2019lsst}, Javalambre Physics of the Accelerating Universe Astrophysical Survey (J-PAS) \citep{benitez2014j}, Subaru Hyper Suprime-Cam (HSC) \citep{hikage2019cosmology} and Physics of the Accelerating Universe (PAU) Survey \citep{serrano2023physics}, it is important to develop the literature with quantitative analyses of this type of correlation.
Furthermore, the use of the angular power spectrum (APS; $C_\ell$) is especially well suited for these analyses, as it naturally incorporates all the information within a given redshift bin and remains robust against photometric redshift (photo-z) errors, in contrast with three-dimensional $P(k)$ approaches that require accurate radial positions.

In this paper, we focus on the use of synthetic data in the harmonic space, to investigate the synergy between future observations conducted by the BINGO radio telescope and the LSST photometric galaxy survey.
We chose the LSST survey as our case study, mainly motivated by its large overlap in area (the entire BINGO sky coverage, $\sim 5000 \textrm{deg}^2$) and redshift with BINGO. 
Notice that, unlike \cite{cunnington2019impact}, where the galaxy sample is selected by matching the redshift range of the \hi survey, we consider the entire photo-z bin, that is, a broader redshift distribution of galaxies. 
To our knowledge, this approach has only been partially explored through Fisher matrix forecasts, as in \cite{2019/padmanabhan}. Here, we assess the detectability of the 21\,cm signal through the cross-correlation of future BINGO and LSST observations, and investigate the impact of photo-z uncertainties, as well as the presence of foreground contaminants, on such analyses.

This paper is organized as follows: 
Section \ref{SecAPS} present the theoretical modeling of the APS for auto and cross-correlation and how they are measured from simulations; Section \ref{SecInstrument} describes our synthetic data sets, simulating BINGO and LSST surveys; Section \ref{SecAPSestimation} presents the foreground cleaning method applied to BINGO simulations and evaluates the statistical significance of an \hi detection through auto and cross-correlation; studies of how well astrophysical parameters describing the \hi signal would be constrained are summarized in Section \ref{SecParams}; we conclude in Section \ref{SecConclusions}.

\section{Angular power spectrum: theoretical modeling and measurement}\label{SecAPS}
%
%
%

% In this section, we present the theoretical background of the angular power spectrum (APS) modeling, and describe how it is estimated from the simulated 21\,cm IM maps.
% Following previous discussions from the literature \citep{Scharf1992, Huterer2001, Padmanabhan2007, Thomas2011, Asorey2012, mcleod2017joint, loureiro2019cosmological, shi2020hir4, cunnington2023h}, we start presenting our theoretical framework connecting the underlying matter field to observable quantities. 

%
%
\subsection{Auto- and cross-correlation modeling}\label{SubsecAPSmodeling}

In this work, auto-correlation always refers to H{\sc i}-H{\sc i} correlations, and cross-correlations always refer to H{\sc i}-Galaxies correlations. 
Let $\delta_{X}(\mathbf{k})$ denote the Fourier transform of observed overdensity fluctuation of a tracer $X$ of the matter field inside some volume of Universe. 
It can be modeled in terms of the underlying dark matter distribution at $z=0$, $\delta(\mathbf{k},0)$, as a function of the $\mathbf{k}$ modes and redshift $z$ \citep{loureiro2019cosmological},
\begin{equation}
\label{eq:overdensity_pk}
    \delta_{X}(\mathbf{k},z)\,=\,b_{X}(\mathbf{k},z)D(z)\delta(\mathbf{k},0),
\end{equation}  
where $b_{X}$ the bias term and $D(z)$ is the growth function. 
The power spectrum, $P_{XY}(k,z)$, is such that,  
\begin{equation}
\label{eq:pkx}
    \langle\delta_X(\mathbf{k},z)\delta_{Y}^{*}(\mathbf{k}^{\prime},z) \rangle\,=\,(2\pi)^3 \,\, \delta^D(\mathbf{k}-\mathbf{k}^{\prime}) \,\,  P_{XY}(k,z),
\end{equation}  
where $\delta^D$ is the Dirac delta function. 
The indexes $X,Y$ represent the biased matter tracers, with $X = Y$ for the auto-power spectrum, and $X \neq Y$, for the cross power spectrum.  
It can be written as a function of the matter power spectrum at $z=0$, $P(k)$, as $P_{XY}(k,z)\,=\,b_X \, b_Y \, r \, D^2(z) \, P(k)$, where $r$ is the correlation coefficient, accounting for the stochasticity between the two fields \citep{wolz2016intensity, Wolz_gbt:2022, cunnington2023h}.

In this work, $X =$ \hi and $Y = g$ (galaxy survey). 
Projecting their three-dimensional distributions (Equation \ref{eq:overdensity_pk}) on the sky by integrating them along the line of sight $\mathbf{\hat{n}}$, leads to
\begin{subequations}
\begin{align}
  \delta_g(\mathbf{\hat{n}})\,&=\, \int dz \delta_g(\chi(z)\, \mathbf{\hat{n}},z)n(z),\\
  \delta_{\text{\hi}}(\mathbf{\hat{n}})\,&=\, \int dz \delta_{\text{\hi}}(\chi(z)\, \mathbf{\hat{n}},z)\phi(z) \Bar{T}_{HI},
\end{align}
\end{subequations}
where $\chi(z)$ is the commoving distance to redshift $z$, $n(z)$ is the galaxy normalized redshift distribution, $\phi(z)$ is the projection kernel for \hi IM\footnote{We consider a top-hat kernel where, for a BINGO redshift bin $i$, $\phi(z) = 1/(z_{i, max} - z_{i, min})$ for $z_{i,min} < z < z_{i,max}$ and $\phi(z) = 0$ otherwise.}. 
The average brightness temperature of the 21\,cm emission can be written as $\bar{T}_{\text{\hi}} = 188 \, h \, \Omega_{\text{\hi}}(z) \, [(1+z)^2 / E(z)]$ mK \citep{battye2013h}, where $E(z) = H(z)/H_0$ is the normalized Hubble parameter, $H_0 = 100 \, h \, {\rm km \, s}^{-1} \, {\rm Mpc}^{-1}$, and $\Omega_{\text{\hi}}$ is the \hi dimensionless density parameter.

% A natural basis to deal with  
Decomposing the projected signal in spherical harmonics \citep{dodelson2003modern}, we have
\begin{equation}
    \delta_{X}(\mathbf{\hat{n}})\,=\,\sum_{\ell=0}^{\infty}\sum_{m=-\ell}^{\ell} a_{\ell m} Y_{\ell m},
\end{equation}
where $a_{\ell m}$ are the coefficients and $Y_{\ell m}$ the harmonic functions. 
The $a_{\ell m}$'s are related to the overdensity through $a_{\ell m}^{X}\,=\,\int d\Omega \delta_{X}(\mathbf{\hat{n}})Y_{\ell m}(\mathbf{\hat{n}})$. Following \cite{loureiro2019cosmological, 2011/sobreira}, the coefficients can be written as
\begin{equation}
\label{eq:alm}
    a_{\ell m}^{X}\,=\,\frac{4\pi \,\, i^l}{(2\pi)^3} \,\, \int d^3k \,\, W^i_{X,\ell}(k) \,\, \delta(\mathbf{k},0) \,\, Y_{\ell m}(\mathbf{\hat{k}}), 
\end{equation}
with the redshift dependence gathered in the window function
\begin{equation} \label{eq:windows}
  W^i_{X,\ell}(k)\, =\, \int dz \,\, b_X(z) \,\, \phi'(z) \,\, D(z) j_{\ell}(k\chi(z))
\end{equation}
where $\phi'(z) = n(z)$, for $X=g$, and $\phi'(z) = \phi(z) \,\, \bar{T}_{\text{\hi}}$, for $X=\text{\hi}$; $j_{\ell}$ is the spherical Bessel function. 
Therefore, the APS can be written as a function of the matter power spectrum
\begin{equation}\label{eq:clls_theo}
    \begin{split}
        C_{XY, \ell}^{ij} &\equiv \langle a_{X, \ell m}^{i} \, a_{Y, \ell m}^{j\,*}  \rangle %\label{eq:cls_alms}
        \\
         &=\, \int dk \,\, k^2 \,\, W_{X,\ell}^i(k) \,\, W_{Y,\ell}^j(k) \,\, P(k), 
    \end{split}
\end{equation}
where the indices $i,\,j$ denote the tomographic redshift bins. 

The theoretical APS used in this paper, for auto- ($i=j$ and $X=Y$) and cross-correlations ($i\neq j$, for the same tracer, $X = Y$, or for different tracers, $X \neq Y$), were computed using the Unified Cosmological Library for $C_{\ell}$ (\texttt{UCLC$\ell$}) code \citep{mcleod2017joint, loureiro2019cosmological}. 
\texttt{UCLC$\ell$} calculates the APS by integrating the \hi projection kernel $\phi(z)$ and the galaxy redshift distribution $n(z)$, as described in Equations \ref{eq:windows} and \ref{eq:clls_theo}. 
It employs the clustering information from the model power spectrum $P(k)$ as computed with the {\tt CLASS} Boltzmann code \citep{2011/lesgourgues, 2011/blas}. 
We account for the non-linear evolution of the Universe as well as the impact of redshift-space distortion effect, implemented in the {\tt UCLC$\ell$} as described in details by \cite{loureiro2019cosmological}. 

\subsection{Angular power spectrum calculation}

In full-sky analyses, the APS can be calculated by expanding the projected cosmological fields into spherical harmonics and using Equation \ref{eq:clls_theo}. 
The partial sky coverage, on the other hand, requires taking into account the coupling of different multipoles as introduced by the geometry of the considered region. 
We employ the pseudo-$C_\ell$ formalism as implemented in the {\tt NaMaster} software \citep{alonso2019unified}. 
It relates the observed APS, $\hat{C}_\ell$, and the true one, $C_\ell$, as $\hat{C}_\ell = \sum_{\ell'} M_{\ell \ell'} C_{\ell'}$, where mode-coupling matrix, $M_{\ell \ell'}$, depends only on the geometry of the considered sky area \citep{hivon2002master}. 
After analytically calculating this matrix, it is inverted to estimate the $C_{\ell}$.

Here, we measure the pseudo-$C_\ell$'s, for auto- and cross-correlation, for 51 linearly spaced multipole bands with width $\Delta \ell = 15$, to guarantee the mode-coupling matrix to be invertible. %\gabriel{The multipoles will running from 24 to 534, but will be limited in future analyses.}

\section{Synthetic data}\label{SecInstrument}

In this section, we describe the production of synthetic sky maps for both tracers considered here, namely,
% describe the methods used to simulate the signal from both datasets, 
the BINGO-like 21\,cm IM and the LSST-like photometric galaxy catalogs. 
We follow closely the methodology described in \cite{mericia2023testing, novaes2022bao, 2023/Novaes_ML} for the BINGO-like simulations and in \cite{zhang2022transitioning, hikage2019cosmology,LSST:2018,LSSTHandBook2011} for the LSST-like simulations.
All simulations are produced using {\tt HEALPix} pixelization scheme \citep{2005/gorski}, with the common resolution of ${\rm N}_{\rm side} = 256$. 

% To simulate both cosmological signals, \hi brightness temperature and galaxy distribution, 
We use the Full-sky Log-normal Astro-fields Simulation Kit\footnote{\href{http://www.astro.iag.usp.br/~flask/}{http://www.astro.iag.usp.br/~flask/}} \citep[{\tt FLASK};][]{xavier2016flask}, which generates lognormal realizations of the sky in tomographic redshift bins ($z$-bins). %, which are particularly well suited for the proposed study because they preserve the non-Gaussian features of the density field and ensure that both tracers trace the same underlying large-scale structures, as expected in real observations, enabling a meaningful cross-correlation signal. 
%
% The 30 BINGO redshift bins and the 3 LSST photometric bins are schematically represented on the Figure \ref{FigZbins}. 
{\tt FLASK} uses as input the theoretical auto- and cross-APS, $C^{ij}_{XY}(\ell)$, of the cosmological fields (all possible combinations of different tracers and redshift bins) 
accounting for their respective projection kernels (selection functions), $\phi(z)$, for the 21\,cm IM, and $n(z)$, for the galaxy number count distribution (see section \ref{SubsecAPSmodeling}).
These realizations mimic the statistical properties of these matter tracer fields, such as their clustering and log-normal distribution. The theoretical $C^{ij}_{XY}(\ell)$ 
employed here were calculated using the {\tt UCLC$\ell$} code. 
We follow companion papers \citep{novaes2022bao, zhang2022bingoVI} and assume our fiducial cosmological model as given by the WMAP 5-year results \citep{dunkley2009five}, $\Omega_m = 0.26$, $\Omega_b = 0.044$, $\Omega_\Lambda = 0.74$, and $H_0 = 71 ~{\rm km}~{\rm s}^{-1}~{\rm Mpc}^{-1}$. 
%We emphasize that our analyses and conclusions do not rely on the assumed cosmology.
% anisotropies.

%Furthermore, the addition of 
In the case of 21\,cm IM, our semi-realistic simulations also account for observational and instrumental aspects, such as the sky coverage, instrumental thermal noise and the presence of foreground contaminants, as we describe bellow. %(as well as its remotion through the foreground cleaning process) account for the specific characteristics of the BINGO instrument is an important step to add realism to the cross-correlation signal, as described on the following section.

\subsection{\hi intensity mapping simulation}
%
%
%

% In this section, we present the simulation pipeline used to model \hi IM observations with the BINGO experiment. 
% Our setup includes \hi cosmological mock maps, instrumental noise, and a variety of astrophysical foregrounds, allowing a realistic environment to test component separation techniques and assess the detectability of the 21\,cm signal.

%
% Esta seção pode seguir o que tem no artigo de bao, já que as simulações foram construídas exatamente da mesma forma. Precisa explicar a binagem em freq (redshift) do bingo, o feixe, a cobertura do céu, o ruído instrumental e os foregrounds que foram incluídos.
% 
\subsubsection{Cosmological signal} \label{Sec:hi_sims}

% The cosmological \hi signal is simulated using the {\tt FLASK}\footnote{\href{http://www.astro.iag.usp.br/~flask/}{http://www.astro.iag.usp.br/~flask/}} code \citep{xavier2016flask}, which generates fast full-sky lognormal realizations. 
A total of $1500$ independent realizations of the projected cosmological \hi signal were produced using {\tt FLASK}. 
We consider the redshift range $0.127<z<0.449$, which corresponds to the frequency %\footnote{From the Doppler law, $\nu_\text{emitted}/ \nu_\text{observed} = 1 + z$, where $\nu_\text{emitted} \approx 1420$ MHz is the rest frame emission frequency for \hi \citep{pritchard201221cm}.} 
interval $980-1260$ MHz, in which the BINGO radio telescope will operate.
This range is divided into 30 uniform frequency bands, each with a bandwidth $\Delta \nu = 9.33 \, \text{MHz}$, corresponding to $z$-bins width of $\Delta z$ varying from $\sim0.010$ to $0.015$ %inside the BINGO redshift range
for low to high $z$-bins. 
The values of $\Omega_{\text{\hi}}$ and $b_{\text{\hi}}$ in each \hi $z$-bin ranges from $3.03 \times 10^{-4}$ to $2.81 \times 10^{-4}$ and from $1.15$ to $1.01$, respectively, decreasing with redshift \cite{zhang2022bingoVI}.
% The bins are modeled as ``top-hat'' functions in redshift space, simplifying the selection function for simulations. 
% The sky maps are produced in the {\tt HEALPix} pixelization scheme \citep{2005/gorski}, with Nside $= 256$, which matches the angular resolution of BINGO and allows for the capture of large-scale features relevant to \hi intensity mapping. 
% The maps are convolved with a Gaussian beam of $\theta_{FWHM} = 40$ arcminutes, corresponding to the main lobe of the BINGO telescope and a multipole $\ell \approx 406$ on harmonic space of APS analyses.
%
We chose to use here the same number of channels used in previous BINGO papers  \citep{liccardo2022bingo, fornazier2021bingoV, zhang2022bingoVI, novaes2022bao, 2023/Novaes_ML} and we will identify them with numbers from 1 to 30 with increasing redshift. %%% Muito bom já citar aqui.
However, it is worth reminding that BINGO hardware allow other choices for the number of frequency bands. This number will be defined taking into account what is most suitable for cosmological analyses, as well as for
% This configuration reflects an initial choice that aligns with the capabilities of the BINGO hardware and data analysis pipeline. 
% However, the channel configuration remains flexible, allowing adjustments to optimize cosmological analyzes or improve the efficacy of 
the foreground cleaning process \citep{mericia2023testing}.

\subsubsection{Sky coverage, instrumental effects and foreground signals}\label{SubsecForegroundsInstruments}

BINGO is designed to survey a sky area of 5324 square degrees (sky fraction of $f_{sky} = 0.129$),
%in the southern sky. Its optical setup allows for a 
in sky transit mode, covering  
%using fixed dishes to scan 
a strip of $\Delta \delta = 14.75^\circ$ wide at declination centered in $\delta = -15^\circ$. %, spanning from $-22.5^\circ$ to $-7.5^\circ$. 
% {\gabriel{The simulated maps are convolved with a Gaussian beam of $\theta_{FWHM} = 40$ arcminutes, corresponding to the main lobe of the BINGO telescope and a multipole $\ell \approx 406$ on harmonic space.}}
To minimize contamination of astrophysical sources from the Galactic plane, %as it introduces significant residual contamination in the final maps, this region was excluded from the current simulations using an apodized mask 
we use a mask reproducing the BINGO footprint, including a Galactic cut to remove strong foreground emission \citep[see][]{mericia2023testing, novaes2022bao}. 
This mask is apodized with a 5 deg cosine transition using the {\tt NaMaster}\footnote{\href{https://namaster.readthedocs.io/en/latest/}{https://namaster.readthedocs.io/en/latest/}} code \citep{alonso2019unified}, avoiding the impact of sharp edges. 
%, which selects only the part of the sky covered by BINGO without the impact of sharp edges in the masked region.

The foreground emissions from Galactic and extragalactic sources are the primary contaminant to the 21\,cm intensity mapping signal. The total observed signal combines the foregrounds, the cosmological \hi signal, and instrumental noise \citep{battye2013h},
\begin{equation}
\label{EqTbrightness}
    T_{\text{obs}}(\nu, \mathbf{\hat{n}}) = T_{\text{\hi}}(\nu, \mathbf{\hat{n}}) + T_{\text{foregrounds}}(\nu, \mathbf{\hat{n}}) + T_{\text{noise}}(\nu, \mathbf{\hat{n}}).
\end{equation}
Following \cite{fornazier2021bingoV, mericia2023testing, novaes2022bao, 2023/Novaes_ML}, we employ the \textit{Planck} Sky Model software \citep[PSM;][]{delabrouille2013pre} to simulate the contribution of the main foreground components in each of the 30 BINGO frequency bands. Our simulations %\alex{include}
include the following components: 
%account for the Galactic synchrotron and free-free emissions, the main foreground \alex{contributions}
%signals contributing to in the BINGO frequency range. \alex{For completeness, we also included}
%,in addition to the Galactic thermal dust, anomalous microwave emissions, the extragalactic contribution from unresolved point sources and the thermal and kinetic Sunyaev-Zeldovich effects. % (see \cite{mericia2023testing, novaes2022bao} for detailed description on how each foreground component is simulated using PSM). } \alex{A detailed description of how each foreground component is simulated in the PSM framwork can be found in \cite{mericia2023testing, novaes2022bao}.}
% {\color{blue} These include: 

%\alex{
\begin{itemize}[leftmargin=*]
    \item \textbf{Galactic Synchrotron Emission:} The dominant foreground component at the BINGO frequencies. The simulation is based on the 408 MHz all-sky map produced by \cite{2015/remazeilles}, extrapolated to the BINGO frequency range using a spatially variable spectral index map derived from \cite{2008/miville-deschenes}. This model accounts for variations in synchrotron spectral properties across the sky.
    \item \textbf{Galactic Free-Free Emission:} Modeled using the H$\alpha$ emission map from \cite{2003/dickinson}. The emission is scaled uniformly across the sky with a slowly varying spectral dependence, providing a smooth foreground template for subtraction.
    %\alex{Esse artigo do Clive é anterior ao WMAP, quase com certeza há referencias mais recentes. A referencia no site lambda.gsfc.nasa.gov é Ade et al. "Planck 2015 results X. Diffuse component separation: Foreground maps". A\&A 594, A10 (2016)     \citep{adam2016planck}}
    \item \textbf{Thermal Dust Emission:} Simulated using templates derived with the {\tt GNILC} method applied to Planck 2015 data \citep{olivari2016extracting}. Dust spectral index and temperature maps are fitted to {\tt GNILC} maps in multiple frequencies and extrapolated to the BINGO bands using a modified blackbody law. %\alex{A\&A 641, A11 (2020) https://doi.org/10.1051/0004-6361/201832618 - Polarized dust foreground ou a referencia de Free-free. \citep{akrami2020planck}}
    \item \textbf{Anomalous Microwave Emission (AME):} %AME is attributed to the rapid rotation of electric dipoles associated with small dust grains. The AME component is
    Simulated using a high-resolution thermal dust template derived from Planck observations. The scaling of AME to low frequencies is based on the ratio between AME and thermal dust found in \cite{planck2016planck}. This component is extrapolated to BINGO frequencies using a single emission law. %\alex{A referencia mais atual desse foreground é Dickinson et al. "The State-of-Play of Anomalous Microwave Emission (AME) research" https://doi.org/10.1016/j.newar.2018.02.001 \citep{dickinson2018state}}
    \item \textbf{Extragalactic Point Sources:} Modeled using %observational 
    point source catalogs at 850 MHz and 4.85 GHz \citep{planck2016planck}. %\textcolor{red}{Que catálogos são esses? O referee vai cobrar essa referencia. Se a referencia do Planck 2015 que eu mencionei antes e está comentada no arquivo .tex, ela tem que aparecer no começo. Se nao for, inclua a referencia,} 
    A population of sources below the detection threshold is simulated based on theoretical number counts, ensuring the maps accurately reflect the flux distribution of faint extragalactic sources. %\alex{    De onde vem esses dados? O Planck produziu um catalogo de fontes pontuais extragalacticas em irsa.ipac.caltech.edu/data/Planck/html/pccs\_hfi\_857\_dd.html. Uma referencia atual - e acho que a mais completa - é Ade et al. "Planck 2015 results XXVI. The Second Planck Catalogue of Compact Sources". A\&A 594, A26 (2016) \citep{planck2016planck}.}
    \item \textbf{Sunyaev-Zeldovich (SZ) Effects:} %The thermal and kinetic SZ effects are extragalactic contaminants, 
    Modeled using the number density as a function of mass from existing galaxy clusters, while the redshift is predicted following \cite{tinker2010large}. %  by galaxy clusters, are also included. While their contribution is less significant at BINGO frequencies, they are simulated for completeness.
\end{itemize}
%}
%\cpn{[\textit{GABRIEL: veja me comentário nessa parte! Por favor, trabalhe um pouco mais nessa parte.}]}

After including the foreground contaminants, the simulated maps are convolved with the telescope beam, approximated here by a Gaussian function with a full width at half maximum (FWHM) of $\theta_{FWHM} = 40$ arcmin for all frequency bands \citep[see][for a complete description of the BINGO beam]{wuensche2022bingo}.  %, corresponding to the main lobe of the BINGO telescope and a multipole $\ell \approx 406$ on harmonic space.
Finally, we add thermal (white) noise, generated as described in \cite{fornazier2021bingoV}. %is an inherent factor in the observations, directly influenced by the optical design and scanning strategy of BINGO. 
The noise level per pixel is estimated considering a system temperature of $70$ K, %the frequency band is equally divided in $30$ bins with channel bandwidth of $\Delta \nu = 9.33 \, \text{MHz}$, 
with $5$ years of observation and a set of $28$ horns, assuming the arrangement design as planned for Phase 1 of the BINGO telescope observation \citep{wuensche2022bingo, abdalla2022bingo3, liccardo2022bingo}. 
From the expected noise level per pixel (the same for all frequencies; see \cite{mericia2023testing, fornazier2021bingoV} for details), % \gabriel{as done in \cite{mericia2023testing}}), 
we can produce as many realizations of thermal noise maps as needed. % modeled using the radiometer equation \citep{2015/bigot-sazy}:
%
%\begin{equation}
%    \sigma_{\text{th}} = \frac{T_{\text{sys}}}{\sqrt{\Delta \nu t_{\text{pix}}}},
%\end{equation}
%
%where $T_{\text{sys}}$ is the system temperature, set at 70 K \citep{wuensche2022bingo}, $\Delta \nu = 9.33 \, \text{MHz}$ is the channel bandwidth, and $t_{\text{pix}}$ is the pixel's integration time. This equation highlights how the observational strategy and system design influence the noise characteristics throughout the survey.

%
\begin{figure}[!ht]
    \centering
    \includegraphics[width=0.5\linewidth]{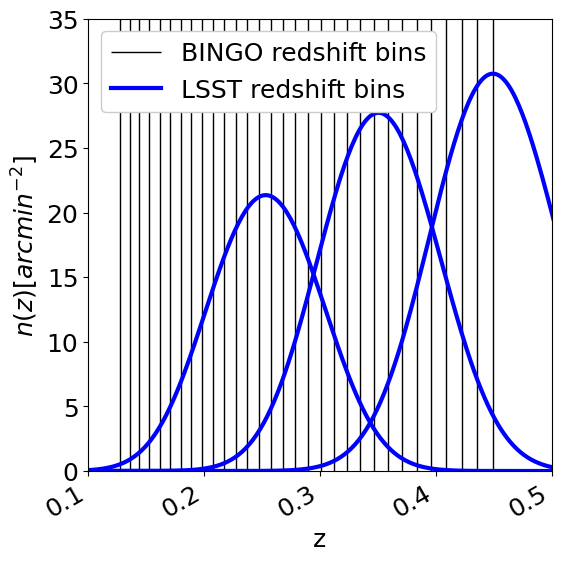}
    \caption{Comparison among BINGO and LSST redshift bins used on the analyses. The black vertical lines delimit the 30 BINGO redshift bins. The blue curves represent the first three LSST photometric redshift bins expected for 10 years of observations (with photo-$z$ error values, $\sigma_Z$, of $0.0375$, $0.0405$ and $0.0435$) \cite{zhang2022transitioning}, chosen given the overlap with BINGO redshift range.}
    \label{FigZbins}
\end{figure}
\subsection{Photometric galaxy catalog simulation} \label{Sec:lsst_sims}
%
%
%

%Along with the 21\,cm simulations, we generate the photometric galaxy distribution maps, following the expected correlation among them as given by the theoretical APS, $C^{ij}_{XY}(\ell)$.
%
% To simulate the photometric galaxy catalog %(Figure \ref{FigRealisticMaps})
% of the \textit{Legacy Survey of Space and Time} (LSST), we employed the {\tt FLASK}. This code enables the generation of log-normal stochastic fields on the sky while maintaining consistency with input statistical functions such as the angular power spectrum $C_\ell$ and selection functions $n(z)$.
%
We adopt the radial selection function for the LSST photometric galaxy catalog as described in \cite{zhang2022transitioning}, parametrized as:
\begin{equation}\label{eq:dNdz}
    \frac{dN}{dz} \propto z^2 \exp\left[-\left(\frac{z}{z_0}\right)^{\beta}\right],
\end{equation}
where $z_0 = 0.24$ and $\beta = 0.90$, taken from \cite{LSST:2018}, %%% Ok!
were derived by fitting to the LSST mock galaxy catalog CATSIM\footnote{https://www.lsst.org/scientists/simulations/catsim} to reproduce the expected performance over 10 years of observations of LSST in terms of depth, photometric redshift quality, and galaxy number density.  % as outlined in the study. 
% This parameterization was essential to ensure that the simulated redshift distribution was representative of the anticipated LSST photometric catalog.

The LSST-like simulations are produced with the {\tt FLASK} package, taking as input the theoretical auto- and cross-APS, as well as the radial selection function (Eq. \ref{eq:dNdz}), in order to generate Poisson sampled lognormal realizations of the projected galaxy number counts in each redshift bin. 
The redshift distribution, $n(z)$, for each tomographic bin is modeled by convolving the radial selection function with a Gaussian distribution centered at the mean redshift $\bar{z}$ and standard deviation given by the photometric redshift error, $\sigma_z = 0.03(1+z)$ \citep{zhang2022transitioning}. 
Here, we consider only the first three redshift bins from the ten bins expected for 10 years of LSST survey, centered at $\Bar{z} = 0.25, 0.35$ and $0.45$ (blue curves in Figure \ref{FigZbins}). 
The photometric redshift errors on each tomographic bin are, respectively, %%% Ok!
$\sigma_z = 0.0375, 0.0405$ and $0.0435$, and the galaxy bias $b_{G} = 1.13, 1.19, 1.25$.

\section{\hi signal detection}
\label{SecAPSestimation}

In this section, we describe the methods employed to recover the 21\,cm cosmological signal from the contaminated maps, presenting the foreground cleaning process and the estimated APS. 
The statistical significance of the auto- and cross-correlation detection are also evaluated.

\subsection{Foreground cleaning}

Recovering the cosmological signal in 21\,cm IM experiments requires an efficient foreground mitigation strategy, since contamination by astrophysical sources can exceed the cosmological signal by several orders of magnitude. 
% {\gabriel{, and foreground separation is achievable because these contaminants exhibit smooth spectral dependence, while the \hi signal shows non-smooth variations due to clustering effects}. 
To address this challenge, we use the Generalized Needlet Internal Linear Combination ({\tt GNILC}) method \citep{Remazeilles:2011}, which rely on the characteristic smooth spectral dependence of these contaminants in contrast to the cosmological signal \citep{battye2013h, alonso2015blind}. 
{\tt GNILC} is a nonparametric component separation technique that uses spatial and spectral information to separate foregrounds from the signal of interest. 
In particular, the method relies on the expected high correlation of the Galactic and extragalactic emissions among the frequency bands, in contrast to a weak correlation of the 21\,cm signal between different bands. 
As the thermal noise has no such correlation, {\tt GNILC} recovers the cosmological \hi IM along with the thermal noise component in each frequency band.  %; a residual foreground contamination is also expected to contribute to the reconstructed map. 
This foreground cleaning method has already been tested on synthetic 21\,cm IM data \citep{olivari2016extracting}, showing good performance in the case of future BINGO observations \citep{fornazier2021bingoV, liccardo2022bingo, mericia2023testing}. 
Here, we use the {\tt GNILC} method following the procedure of \cite{mericia2023testing} for the BINGO case.

As discussed in \cite{mericia2023testing}, a byproduct of the {\tt GNILC} algorithm are the ILC filters constructed for each needlet (%\sout{a type of wavelet defined on sphere} 
a kind of bandpass filter in harmonic space, defined here by the index $k$), 
%\eduardo{\sout{considered}}
input required by the method, and for each pixel of the maps produced by them, in such a way as to %\eduardo{\sout{filter out} subtract}
the foreground contaminants, while preserving the 21\,cm signal and the thermal noise. 
Each ILC filter, ${\bf W}^{(k)}(p)$, is applied to a synthetic observation (cosmological signal and contaminants), $\mathbf{x}^{(k)}(p) = \mathbf{s}^{(k)}(p) + \mathbf{n}^{(k)}(p) + \mathbf{f}^{(k)}(p)$, % obtaining the $\mathbf{\hat{s}} = \mathbf{W x}$ map, 
where $\mathbf{s}^{(k)}(p),  \mathbf{n}^{(k)}(p)$, and $\mathbf{f}^{(k)}(p)$ are the 21\,cm signal, thermal noise, and foreground contamination, respectively, at each pixel $p$ and at each needlet scale $(k)$. 
The set of filtered maps, %\eduardo{\sout{$\mathbf{\hat{s}} = \mathbf{W x}$}} 
$\mathbf{\hat{s}}^{(k)}(p) = \mathbf{W}^{(k)}(p) \mathbf{x}^{(k)}(p)$, for each needlet angular scale $k$, are combined to obtain the reconstructed 21\,cm signal plus noise map at each frequency band.
%
%\eduardo{\sout{In in addition to the noise, $\mathbf{W n}$, a residual foreground contribution, $\mathbf{W f}$, is also expected to contaminate the reconstructed 21\,cm maps, such that,}}
Omitting the indices $p$ and $k$, the reconstructed maps $\mathbf{\hat{s}}$ can be written as
\begin{equation} \label{eq:recHI_plus_residue}
    \mathbf{\hat{s}} = \mathbf{W s} + \mathbf{W n} + \mathbf{W f} \, ,
\end{equation}
%
%\eduardo{\sout{where the ILC filter $\mathbf{W}$ can be used to quantify the contribution of each contaminant.}} 
where the ILC filter \( \mathbf{W} \) can be used to quantify the impact of each %\eduardo{\sout{contaminant}} 
component, such that \( \mathbf{W n} \) accounts for thermal noise, and \( \mathbf{W f} \) corresponds to the residual foreground contamination. Besides the residual contamination, it is also expected that, due to the product $\mathbf{W s}$, a fraction of the cosmological signal $\mathbf{s}$ is lost during the foreground cleaning process (see, for example, \citep{Wolz_gbt:2022, cunnington2023h}), %\gabriel{
which affect the cross-correlation with galaxy surveys, as we discuss in the next sections.%}. 

\paragraph{\bf APS debiasing process:}
To recover the APS from the 21\,cm signal and its correlation with a galaxy sample, %we follow \cite{mericia2023testing, fornazier2021bingoV} and implement a debiasing procedure. 
%\alex{
we  implement a debiasing procedure, in the same fashion as \cite{mericia2023testing, fornazier2021bingoV}. %}
For auto-correlation, this procedure accounts for two steps: subtracting the expected average contribution of the thermal noise, $\hat{C}_\ell^{noise-{\rm ILC}(i)}$ (additive bias), and dividing by a scale-dependent factor, $\hat{S}^i(\ell)$ (multiplicative bias), to account for a suppression of the APS due to the loss of the 21\,cm signal. 

In our case, the estimation of the additive bias  is done by passing 50 noise realizations, for each frequency band, through the ILC filters produced as described previously. We then average the APS of the 50 ILC-reconstructed noise maps to obtain the $\hat{C}_\ell^{noise-{\rm ILC}, i}$. It is important to mention that these 50 noise realizations are different from the noise realizations used to obtain the synthetic observation maps.
%The additive bias is obtained by passing different realizations of white noise maps, for each frequency band $\nu$, through the filter, $\mathbf{W n}$. 
%Here, we use 50 noise realizations, other that added to the synthetic observation, averaging the APS of the noise map reconstructed from each of them, obtaining $\hat{C}_\ell^{noise-{\rm ILC}, \nu}$. 

To obtain $\hat{S}^i(\ell)$, the multiplicative bias for each redshift bin (frequency band), we apply the ILC filter to $50$ realizations %\eduardo{\sout{of the 21\,cm IM signal}
of 21 cm simulated maps, %\eduardo{\sout{$\mathbf{W s}$,}} 
and then average the ratio between the APS from the ILC filtered 21\,cm maps, $C_\ell^{\text{\hi}-{\rm ILC}(i)}$, and from the input ones, $C_\ell^{\text{\hi} (i)}$, that is, $\hat{S}^i(\ell) = \langle C_\ell^{\text{\hi}-{\rm ILC}(i)} / C_\ell^{\text{\hi}(i)} \rangle$. 
The corrected (debiased) APS for the $m$th 21\,cm IM realization at $z$-bin $i$ is given by, 

\begin{equation} 
\tilde{C}_\ell^{\mathtt{GNILC}-deb (i)} |_m = \frac{\hat{C}_\ell^{\mathtt{GNILC} (i)} - \hat{C}_\ell^{noise-{\rm ILC} (i)}}{\hat{S}^i(\ell)} \bigg |_m. 
\end{equation}

\begin{figure}[t]
    \centering
    \includegraphics[width=\linewidth]{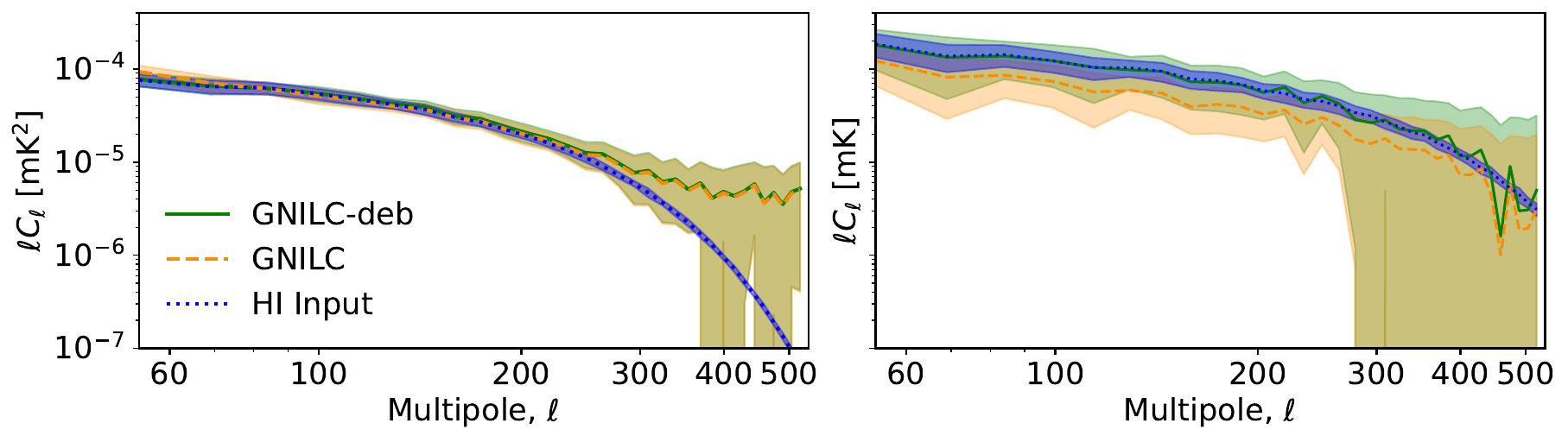}
    \caption{Comparison among the mean APS and 68\% regions from 50 realizations before (GNILC; green solid lines) and after (GNILC-deb; orange dashed lines) applying the debiasing process (see text for details). 
    Results from the input \hi maps are shown in blue (\hi Input; %\gabriel{; 
    blue dot lines). %} 
    Left and right panels correspond to auto-APS of BINGO $z$-bin 14 ($\Bar{z}_{\hi} \approx 0.252$) and its cross-APS with the first LSST photo-z bin ($\Bar{z}_{G} \approx 0.250$), respectively.
    % Left panel: auto-APS; right panel: ``{\hi}xG1'' cross-APS, where G1 is the first LSST photo-$z$ bin. 
    % In both panels, the blue curve shows the APS of the input \hi maps used by GNILC (``HI Input''), the orange curve corresponds to the mean APS obtained from the {\tt GNILC} maps and the green curve shows the APS after applying debias corrections (``GNILC-deb'').
    }
    \label{FigCls_ClsGNILCdebias}
\end{figure}

This correction procedure, originally designed for autocorrelation analyses, is extended here to cross-correlation by setting the additive term to zero, since the white noise contaminating the cosmological signal is not correlated with galaxy maps. 
The multiplicative bias is employed to account for the suppression of the cross-APS due to the 21\,cm signal lost during foreground removal: 
% 
% \begin{equation} \tilde{C}_\ell^{\mathtt{GNILC}-deb (\nu), gal(j)}|_m = \frac{\hat{C}_\ell^{\mathtt{GNILC} (\nu), gal (j)}}{\hat{S}^{\nu, j} (\ell)} \bigg |_m, 
% \end{equation} 
% %
% \eduardo{Outra sugestão seria}
%
\begin{equation} \tilde{C}_\ell^{\mathtt{GNILC}-deb (i), gal(j)}|_m = \frac{\hat{C}_\ell^{\mathtt{GNILC} (i), gal (j)}}{\hat{S}^{i, j} (\ell)} \bigg |_m. 
\end{equation}
%
%\eduardo{\sout{$\hat{S}^{\nu, j} (\ell) = \langle C_\ell^{\text{\hi} (\nu),gal(j)} / C_\ell^{\text{\hi}-{\rm ILC} (\nu),gal(j)} \rangle$}} 
% \eduardo{\sout{$\hat{S}^{i, j} (\ell) = \langle C_\ell^{\text{\hi} (i),gal(j)} / C_\ell^{\text{\hi}-{\rm ILC} (i),gal(j)} \rangle$} 
$\hat{S}^{i, j} (\ell) = \langle  C_\ell^{\text{\hi}-{\rm ILC} (i),gal(j)} / C_\ell^{\text{\hi} (i),gal(j)} \rangle$ is obtained averaging the ratio between the cross-APS among the filtered $\mathbf{s}'$ map and the corresponding galaxy density realization, $C_\ell^{\text{\hi}-{\rm ILC} (i),gal(j)}$, 
and among the original $\mathbf{s}$ map and the same galaxy density map, $C_\ell^{\text{\hi} (i),gal(j)}$. 
Again, the average is obtained using 50 pairs of 21\,cm IM and galaxy density maps. Figure \ref{FigCls_ClsGNILCdebias} presents a comparison of the APS for the two cases.
% This generalization enables us to consistently debias both auto- and cross-correlation power spectra and assess the significance of the recovered cosmological signal.

\paragraph{}
Since the foreground cleaning process is computationally demanding, we apply the {\tt GNILC} method to 50 ($m = $ 1,2, ..., 50) synthetic observations, employing the debiasing process for each resulting auto- and cross-APS. 
Notice that, for each $m$ map, the multiplicative bias is estimated using a different set of 50 \hi IM realizations, not used as the cosmological signal in the synthetic observations. 

In addition to the 50 cleaned BINGO-like maps, we follow \cite{novaes2022bao, 2023/Novaes_ML} and produce a set of 1500 realizations, mimicking these {\tt GNILC} outputs %\gabriel{
(intended to be used to construct the covariance matrix employed in Section \ref{SecParams}). %}. 
For this, we simply add a different realization of the white noise component to each of the 1500 \hi IM realizations, including the expected residual foreground contribution estimated using the ILC filter, $\mathbf{W f}$ (see Equation \ref{eq:recHI_plus_residue}). 
Since we run {\tt GNILC} 50 times (each $m$ map), we use the 50 resulting estimates of the residual foreground contribution to contaminate the 1500 realizations. 

While the white noise is different from one realization to another, the foreground residual is repeated every 50 realizations. 
This way we can produce a larger set of fast BINGO-like simulations, while avoiding to perform the computationally expensive foreground cleaning process for each of them. 
For this set of simulations we do not need the repetition of the two steps debiasing process of the (auto- and cross-) APS, since no cosmological signal is lost. 
For autocorrelation, we perform a simpler debiasing by subtracting the expected constant noise amplitude, $N_\ell$ (the theoretical noise level); cross-APS does not require any debiasing in this case. 

\subsection{Results}
\label{SecResults}
%

% \gabriel{Incluir as análises a respeito do $r$ e fazer a Figura \ref{FigHistContorno} (do histograma), se ela for necessária.}

In this section, we present the results for estimating auto- (HI-HI) and cross- (HI-galaxies) APS in section \ref{SubsecAPS}. 
The impact of thermal noise and foreground residual contamination is individually evaluated. The role of photometric uncertainties in the cross-APS calculations is also %assessed.
quantified. %Finally, we quantify the statistical significance of a potential detection of both auto- and cross-APS signals based on the simulations.
The statistical significance of a potential detection of both auto- and cross-APS signals based on the simulations will be quantified in Section \ref{SecStatsSign}.

\subsubsection{Angular power spectrum estimation}
\label{SubsecAPS}

\begin{figure}[!ht]
    \centering
    \includegraphics[width=\linewidth]{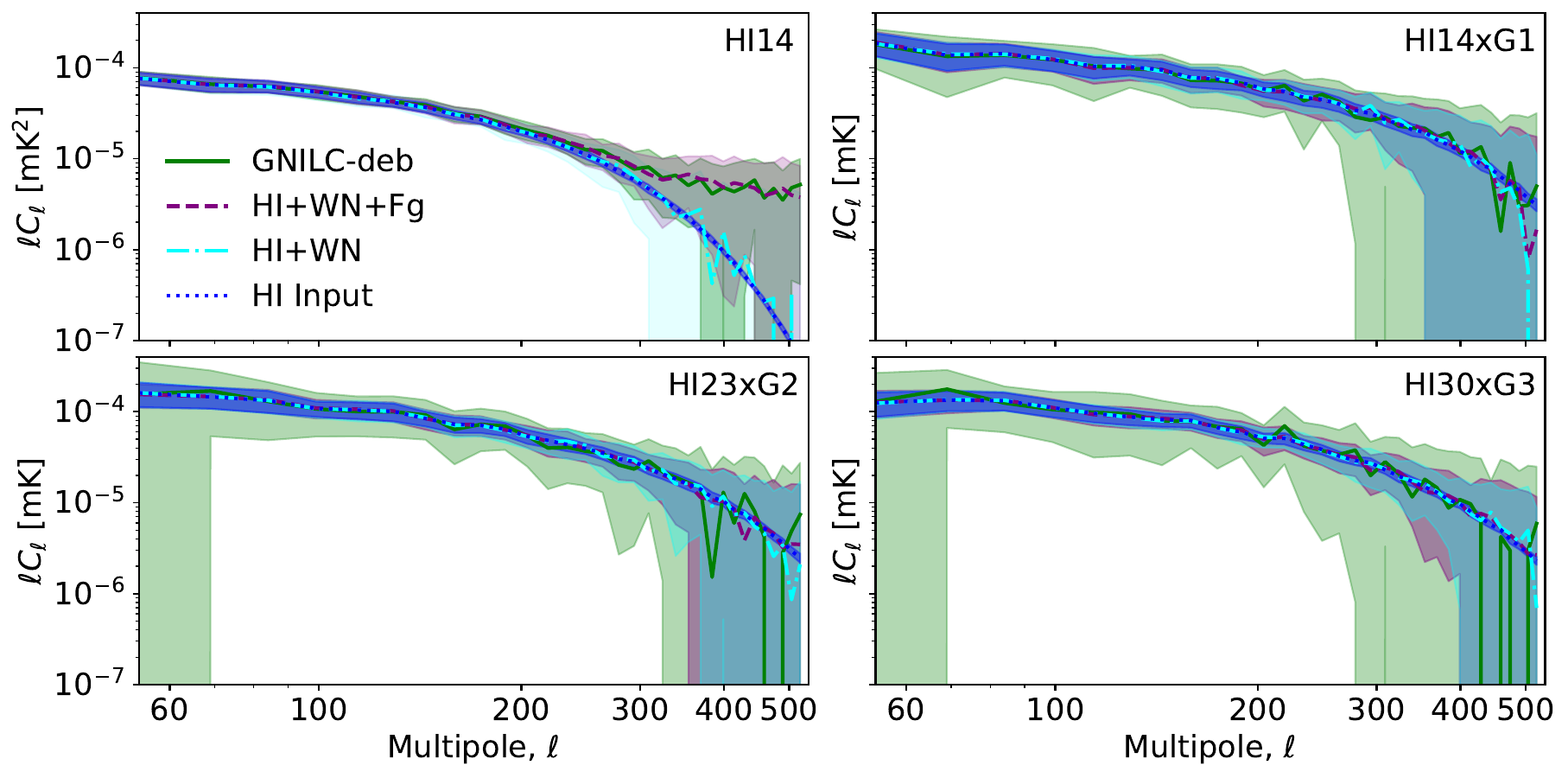}
    \caption{Comparison of the mean APS and 68\% regions from 50 realizations for four different scenarios. Colors and scenarions are discussed in the beginning of section \ref{SubsecAPS}. Upper left panel shows the auto-APS for BINGO z-bin number 14. The other three panels show the cross-APS among BINGO and LSST bins as indicated in the top right corner of each panel (see definitions in Table \ref{Tab_Zbins}). 
    % Upper right, lower left, and lower right correspond to the cross-APS between BINGO z-bin 14, 23, and 30 and LSST bins 1, 2, and 3, respectively (see Table \ref{Tab_Zbins}). %, $\Bar{z} \approx 0.25$.
    % Lower left: cross-APS for LSST bin 2, $\Bar{z} \approx 0.35$.
    % Lower right: cross-APS for LSST bin 3, $\Bar{z} \approx 0.45$.
    } 
    \label{FigCls_ClsAutoCrosHIwnFgGNILCdeb}
\end{figure}

To assess the impact of individual contaminants on the 21\,cm signal, we analyze the angular power spectra of \hi maps under four different scenarios:
\begin{enumerate}[label=(\roman*)]
\item input (original) \hi maps (``HI Input'', blue curve in Figure \ref{FigCls_ClsAutoCrosHIwnFgGNILCdeb}), 
\item \hi maps contaminated with white noise, with the subsequent subtraction of the  expected noise amplitude $N_\ell$ from the auto-APS, %\cpn{
that is, \hi signal contaminated by a residual contribution of white noise %}
(``HI + WN'', cyan curve in Figure \ref{FigCls_ClsAutoCrosHIwnFgGNILCdeb}), 
\item BINGO-like %fast 
simulations, adding the white noise and foreground residuals to the \hi signal, with the subsequent subtraction of the estimated noise amplitude $N_\ell$ from the auto-APS (``HI + WN + Fg'', purple curve in Figure \ref{FigCls_ClsAutoCrosHIwnFgGNILCdeb}), and 
\item \hi signal reconstructed using the {\tt GNILC} cleaning method, with the application of additive (for auto-APS) and multiplicative (for auto- and cross-APS) debiasing corrections (``GNILC-deb'', green curve in Figure \ref{FigCls_ClsAutoCrosHIwnFgGNILCdeb}). 
\end{enumerate}

All of these scenarios account for the instrumental beam effect. We estimate the cross-correlation among the three $z$-bins of LSST-like galaxy density samples and $30$ BINGO-like \hi IM realizations (3$\times$30 cross-APS) for each of these four scenarios, as well as the autocorrelation for all \hi IM $z$-bins.
Unless stated otherwise, we will show, as illustrative examples throughout this section, the results obtained using the three BINGO $z$-bins closest to the center of the LSST redshift distributions (see Fig. \ref{FigZbins}), as summarized in Tab. \ref{Tab_Zbins}. 
As we show later, these are the cases that produce a higher signal-to-noise ratio for the cross-correlation case. 

\begin{table}[t]
    \centering
    \small
    \begin{tabular}{|c|c|c|c|}
         \hline
         BINGO & $\bar{z}_{HI}$ & LSST & $\bar{z}_{G}$ \\
         \hline
         \hi 14 & 0.252 & G1 & $0.250$ \\
         \hi 23 & 0.353 & G2 & $0.350$ \\
         \hi 30 & 0.442 & G3 & $0.450$ \\
         \hline
    \end{tabular}
    \caption{Summary of the photometric $z$-bins and the respective \hi IM $z$-bins centered on them. Numbering the BINGO $z$-bins from 1 to 30, the first and second columns list the number of the three chosen bins and their mean redshift value. The same information is listed for LSST at third and fourth columns, numbering their $z$-bins from 1 to 3.%%% Ótimo.
    }
    \label{Tab_Zbins}
\end{table}

Figure \ref{FigCls_ClsAutoCrosHIwnFgGNILCdeb} shows the mean and 68\% region  (the 1$\sigma$ dispersion of the 50 APS curves) of the APS from 50 realizations, for each of the four scenarios listed above. 
The top panel shows an example of auto-APS for one of the BINGO $z$-bins, from where we observe significantly larger 68\% regions due to the presence of contaminants, in particular at high multipoles (small scales), where the %\alex{
thermal noise amplitude becomes larger than the cosmological signal. %}
%Indeed, it is possible to see that the 
White noise (\hi+WN case) seems to be the main responsible for this behavior, given that the inclusion of foreground residual (\hi+WN+Fg case) and the implementation of the component separation process (GNILC-deb case) results in 68\% regions similar to the \hi+WN case. % \gabriel{Remembering that, given that we do the additive correction, the WN here refers to the white noise residues after the correction.} 
The remaining panels of Figure \ref{FigCls_ClsAutoCrosHIwnFgGNILCdeb} display the results for the cross-APS between three BINGO $z$-bins and each of the LSST $z$-bins (as listed in Tab. \ref{Tab_Zbins}). 
%Comparing the mean curves with those from the first panel, there is a clear excess of power, with respect to \hi Input scenario, in higher multipoles of the auto-APS introduced by the contaminants, which we do not see for cross-APS. 

%\alex{
We also observe a clear power excess in the mean curves of the auto-APS spectra, compared to the \hi Input scenario. The excess appears in the higher multipoles of the auto-APS, introduced by the contaminants, which is not observed in the cross-APS. %}
These results also show that, even though the residual contamination by white noise and foregrounds do not correlate with the galaxy distribution, not contributing to the amplitude of the APS, they boost the noise in cross-correlation, increasing the 68\% region, as expected. 

Similarly to the auto-APS case, white noise seem to be the main driver of the increase of the 68\% region for cross-APS.  
In addition, one can see that the component separation process also impacts the estimate of the cross-APS, contributing with an additional increase of the %\gabriel{
68\% region, likely due to the loss of \hi signal. %}. 
Comparing the first and second panels of Fig. \ref{FigCls_ClsAutoCrosHIwnFgGNILCdeb} (for the same \hi IM $z$-bin), we also notice a wider 68\% region for the cross-APS compared to auto-APS, even when comparing only results from the \hi Input scenario. 
%As we will discuss later in this section, this seems to be associated to the photometric redshift error.
%\alex{
We believe this effect is associated to the photometric redshift error, and will be discussed later in this section.%}

\begin{figure}[t]
    \centering
    % \begin{tabular}{cc}
    %     \includegraphics[width=0.47\columnwidth]{Figuras/SNR_Cls_comparacao_14.pdf} &  \includegraphics[width=0.47\columnwidth]{Figuras/SNR_Cls_comparacao_23.pdf} 
    % \end{tabular}
    
    % \includegraphics[width=0.47\columnwidth]{Figuras/SNR_Cls_comparacao_30.pdf} 
    \includegraphics[width=1\columnwidth]{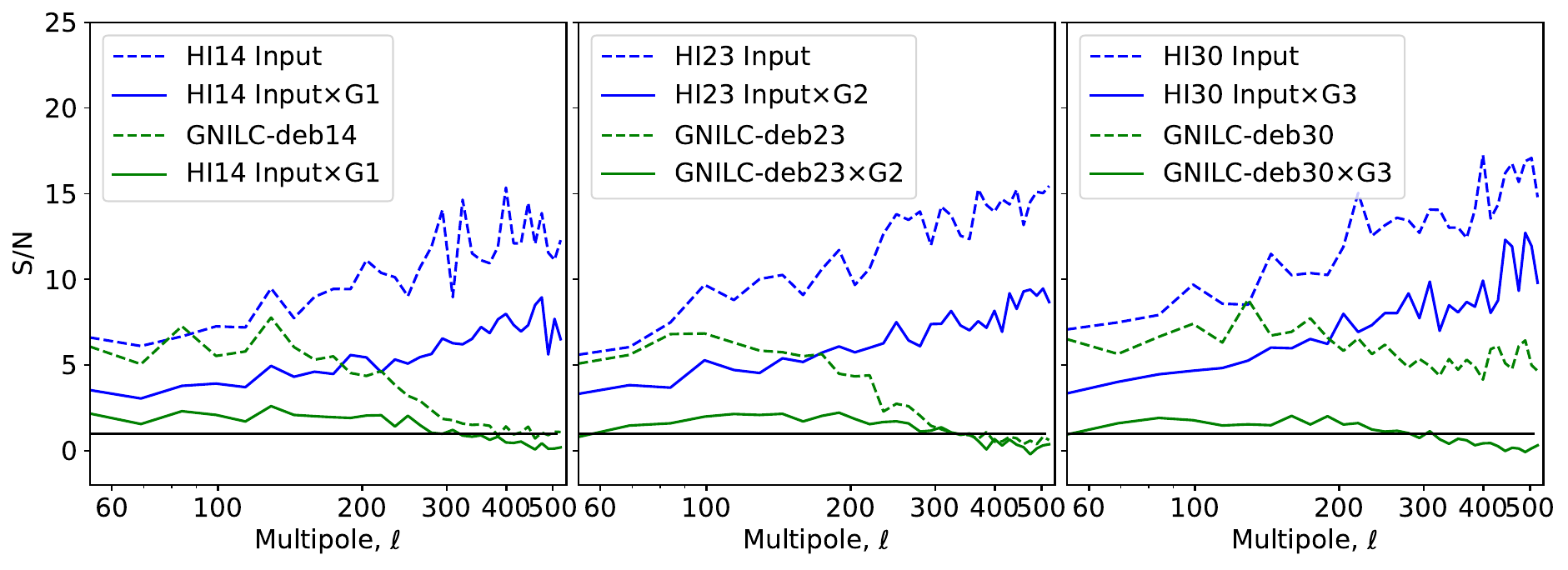}
    \caption{Signal-to-noise ratio (SNR) for \hi IM auto-APS (dashed lines) and their cross-APS with the galaxy density maps (solid lines), averaged over 50 realizations.  % for \hi input maps versus the three photometric bins of LSST. 
    Blue and green lines show results from scenarios \hi Input and GNILC-deb, respectively. BINGO and LSST z-bins are indicated in each panel (see definitions in Table \ref{Tab_Zbins}).
    Black horizontal line represents SNR = 1. 
    % Upper left panel: auto-APS and cross-APS with LSST bin 1 (as shown in Figure \ref{FigZbins}), $\Bar{z} \approx 0.25$.
    % Upper right panel: auto-APS and cross-APS with LSST bin 2, $\Bar{z} \approx 0.35$.
    % Lower panel: auto-APS and cross-APS with LSST bin 3, $\Bar{z} \approx 0.45$.
    }
    \label{FigSNRClsComparingAutoCros}
\end{figure}

Before investigating the impact of photo-$z$ errors, we evaluate the SNR of auto and cross-APS, presented in Figure \ref{FigSNRClsComparingAutoCros}. 
Comparing blue and green curves, we can see, on both the auto- and cross-APS, the impact of foreground residuals and thermal noise remaining in the \hi IM maps after the component separation and debiasing processes, in particular for higher multipoles ($\ell \gtrsim 300$). 
These results also indicate that the cross-APS is more impacted by cleaning process and residual contamination, with a decrease in the SNR over the entire multipole range, while the auto-APS does not show significant decrease for $\ell \lesssim 100$. 
In both cases, the SNR becomes smaller than 1 for $\ell \gtrsim 300$, except in the case of the auto-APS from the last redshift bin (last panel of Fig. \ref{FigSNRClsComparingAutoCros}). 

It is worth noticing from Fig. \ref{FigCls_ClsAutoCrosHIwnFgGNILCdeb} that the amplitude of the auto-APS of the \hi IM $z$-bin number 14 (first panel) is lower than that from its cross-correlation with the first $z$-bin of the galaxy distribution (second panel). 
Still, the first panel of Fig. \ref{FigSNRClsComparingAutoCros} shows the SNR to be higher for its autocorrelation, even for the \hi Input scenario. 
This behavior is a consequence of the larger 68\% region % (1$\sigma$ dispersion) \gabriel{1$\sigma$} 
for cross-correlation, showing its impact on the cross-correlation detection in the future BINGO and LSST surveys. 
As we mentioned before, this larger dispersion might be associated with the photometric redshift uncertainty, $\sigma_z$, a suspicion we explore through the null test described below.

\hi IM instruments survey the Universe using a tomographic approach, where each volume "slice" corresponds to very narrow frequency bands, consequently to a high radial resolution. Current galaxy photometric surveys use filters that, by construction, covers a much larger redshift interval with consequent poorer redshift resolution, as shown in Figure \ref{FigZbins}.
The poor photo-$z$ accuracy can lead to galaxies being assigned to a bin in which they are not physically present. 
In this work, we computed cross-correlations of BINGO-like \hi maps 
%\alex{
%é bom denominar que são BINGO-like maps}
with each of the three LSST $z$-bins using their entire photometric galaxy distribution, instead of slicing them into 30 narrow $z$-bins with widths matching the BINGO bins, as described in \cite{cunnington2019impact}. 
%As opposed 
%\alex{
Contrary %}
to their results, we did not find %\cpn{
a degradation in power %} 
of the cross-APS (see Fig. \ref{FigCls_ClsAutoCrosHIwnFgGNILCdeb}). 

As a safety check, we implement a null test to investigate the impact of the presence of galaxies in the photo-$z$ bin that do not contribute to the \hi signal with which it is being cross-correlated  %\cpn{
(non-correlated galaxies). %}.
We estimate the cross-correlation between a fixed \hi IM realization and %\gabriel{(not spatial statistically correlated)} 
500 maps generated as Gaussian random distributions with zero mean and standard deviation $\sigma$. 
%\cpn{
Given that there is no spatial correlation among them, the full Gaussian distribution mimics the presence of non-correlated galaxies. %}
We repeat this procedure for four different values of $\sigma$, namely $0.0110, 0.0375, 0.0405$, and $0.0435$, each of them representing the width (error) of the redshift bin. 
%Notice that the last three values match the photometric redshift uncertainty (see Sec. \ref{Sec:lsst_sims}), while the former one matches the width of a \hi IM z-bin (see Sec. \ref{Sec:hi_sims}).
%\alex{
Notice that the former $\sigma$ value matches the width of a \hi IM z-bin, while the other three values match the photometric redshift uncertainty, as described in Secs. \ref{Sec:hi_sims} and \ref{Sec:lsst_sims}).%}

\begin{figure}[t]
    \centering
    \includegraphics[width=0.65\linewidth]{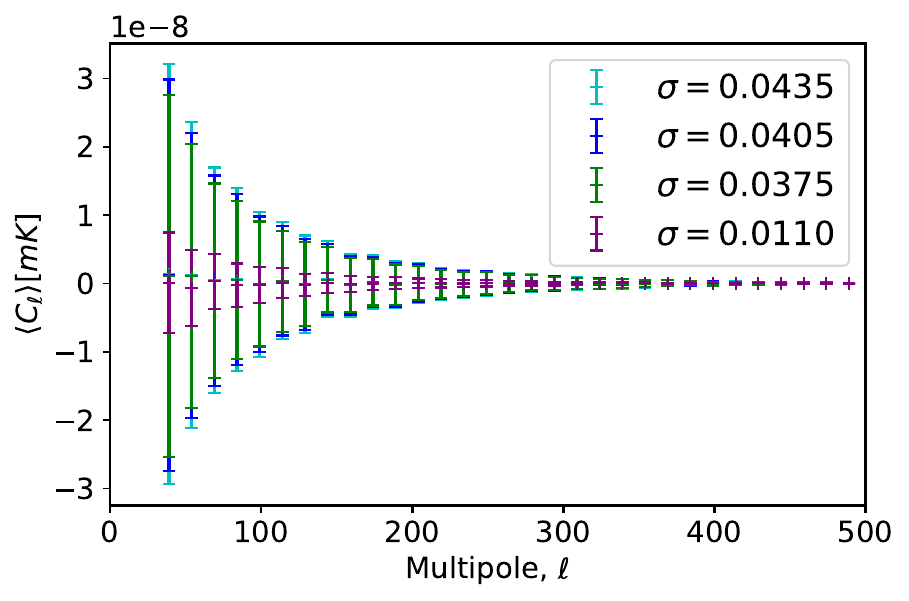}
    \caption{Null test: Average and 68\% error bars from the cross-correlation among a synthetic \hi IM map and 500 %\cpn{
    (non-correlated) %} 
    maps generated as Gaussian random distributions with zero mean and standard deviation of $\sigma$. The results are shown for four values of $\sigma$, representing the width of the photometric redshift distribution of the galaxies (cyan, blue, and green) and the \hi IM $z$-bin (red). 
    }
    \label{FigNullTest}
\end{figure}

Figure \ref{FigNullTest} show the average %\cpn{
(expected to be close to zero) %}
and 68\% error bars over the 500 cross-correlation estimates, for each $\sigma$. 
As we can see, although the $\sigma$ value of the Gaussian random distributions does not introduce any bias in amplitude, it impacts the statistical uncertainty on the APS estimates, leading to larger 68\% errors bars the larger the $\sigma$ value, affecting the entire range of multipoles.
This noise-like behavior explains why the autocorrelation presents higher SNR compared to cross-correlation (Fig. \ref{FigSNRClsComparingAutoCros}), even with the fact that optical galaxy data and \hi IM contaminants are uncorrelated. 
Therefore, cross-correlating \hi IM with the entire photometric galaxy distribution preserves the cross-APS amplitude, while introducing a higher uncertainty in the estimated signal (given by the large 68\% regions). 
Moreover, the null test confirms that the presence of thermal noise in the \hi IM maps increases significantly the 68\% regions at high multipoles, where it starts to dominate the cosmological signal, although not affecting the amplitude of the cross-APS (see Fig. \ref{FigCls_ClsAutoCrosHIwnFgGNILCdeb}).

\begin{figure*}[!ht]
    \centering
    \includegraphics[width=\linewidth]{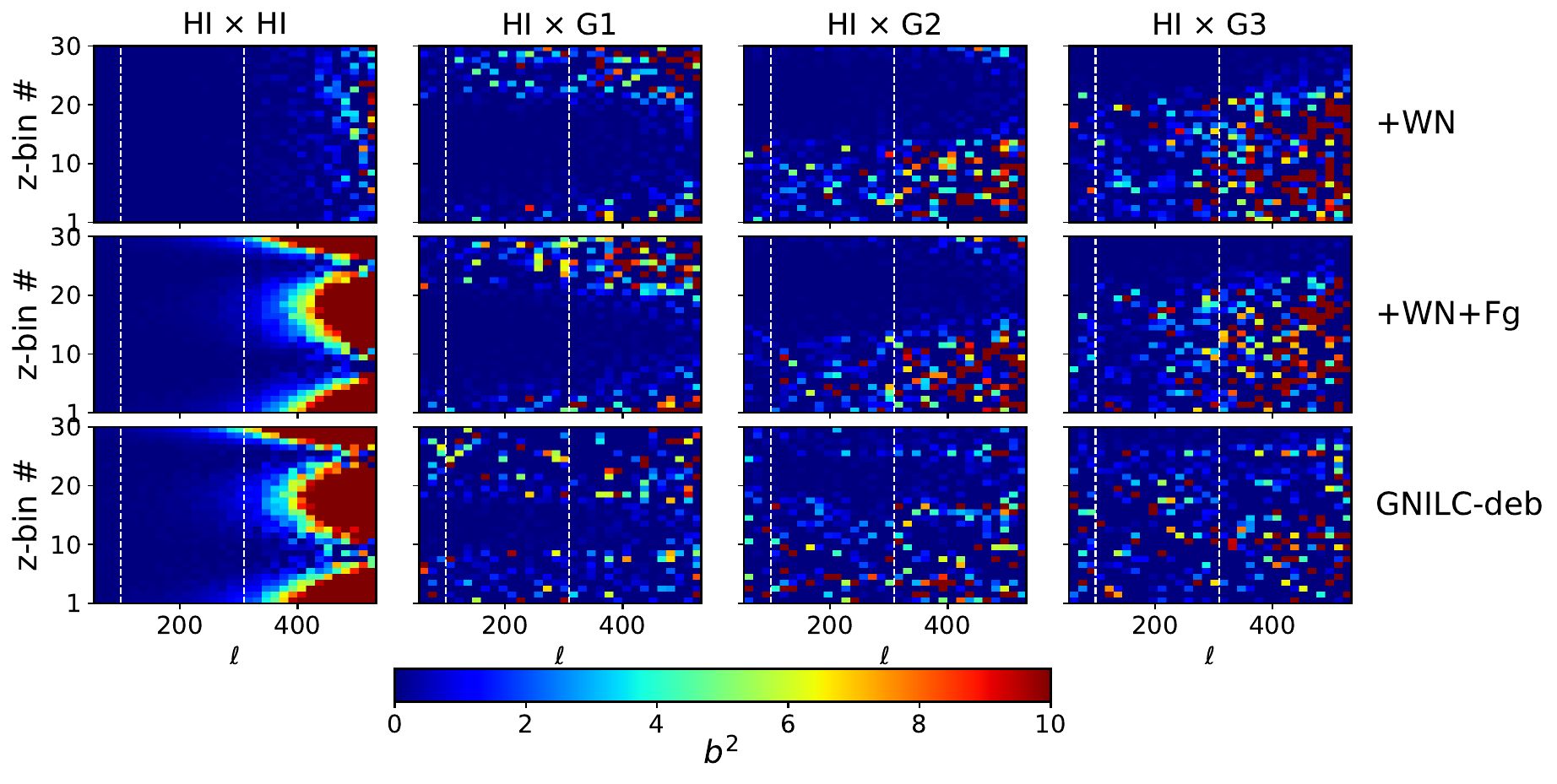}
    \caption{Impact of the white noise, foreground residues, and component separation process on the auto- and cross-APS.
    The panels show the relative difference between the contaminated and clean angular power spectra (Eq. \ref{eq:b2}) for the 30 BINGO redshift bins (y-axes). Top row corresponds to contamination by white noise (HI+WN), middle row describes contamination by white noise and foreground residues (HI+WN+Fg), and bottom row is the GNILC-deb result. Columns show autocorrelation (first column) and cross-correlation with the three LSST photometric bins  (other three columns). White dashed lines in $\ell = 99$ and $\ell = 309$ indicates the range in multipoles considered in the parameter estimations (see Section \ref{SecParams}).
    }
    \label{FigBiasNoises}
\end{figure*}
%

% In contrast to methods that rely on slicing the galaxy sample into narrower redshift intervals (e.g., \citealt{cunnington2019impact}), our strategy preserves the large redshift extent over which the \hi and galaxy surveys are correlated, thus maximizing the usable information content despite the inherent photometric redshift uncertainties. 
Finally, we evaluate the impact of each contaminant (white noise and foreground residuals), as well as the component separation process, on the auto- and cross-APS for all BINGO bins in each scenario (ii–iv). 
For this, we use the relative difference between them and the APS from the \hi Input scenario, $C_\ell^{\text{\hi Input}}$, defined as %%% ok, então não vamos chamar de "foreground bias"?
% Figure \ref{FigBiasNoises} displays the so-called \textit{foreground bias}\footnote{We emphasize that this foreground bias should not be confused with the astrophysical bias parameters of \hi or galaxies, nor with the additive or multiplicative debiasing corrections applied to the cleaned maps.}.
%
\begin{equation} \label{eq:b2}
    b^2(\ell) = \bigg\langle \,\frac{C_\ell}{C_\ell^{\text{\hi Input}}} - 1 \,\bigg\rangle,
\end{equation}
as a function of multipole, $\ell$, and averaged over the $50$ realizations. 
Figure \ref{FigBiasNoises} shows the $b^2$ factor for the auto-APS from all $30$ BINGO redshift bins, as well as their cross-APS with each of the three LSST photometric redshift bins.
The first column shows how the white noise impacts the auto-APS at high multipoles and how the contribution of foreground residuals significantly increases the discrepancy with respect to the original \hi Input signal around $\ell \gtrsim 300$. 
The impact varies with the redshift bin, showing a worse recovery of the \hi signal for frequency bands closer to the border of the BINGO frequency range. 
As shown by \cite{mericia2023testing}, using a larger number of frequency bands can improve the performance of the GNILC, around the center and borders (see bottom right panel of their Fig. 8).
% Figure \ref{FigBiasNoises} reveals that foreground contamination affects recovered auto-APS across a wide multipole range, especially at small scales \gabriel{($\ell \gtrsim 300$)}, and that the impact varies depending on the redshift bin considered. 

%A distinct behavior for cross-correlation is observed in Fig. \ref{FigBiasNoises}, which shows a better estimate of the cross-APS for \hi signal $z$-bins closer to the center of the photometric galaxy distribution.

%\alex{
The cross-correlation plots in Fig. \ref{FigBiasNoises} show a better estimate of the cross-APS for \hi signal $z$-bins closer to the center of the photometric galaxy distribution. 
%} % with LSST photometric redshift bins until $\ell \lesssim 400$, particularly near the central redshift of each bin (see table \ref{Tab_Zbins}). 
% This behavior is visually highlighted by the blue regions concentrated around the central redshift of each galaxy photo-$z$ bin. 
%As you move away from the center, the cross-correlation signal weakens and becomes increasingly dominated by contaminants. 
Moving away from the center causes the cross-correlation signal to weaken and become increasingly dominated by contaminants. 
Compared to the impact of white noise (\hi+WN; first row), the addition of foreground residuals further restricts the cross-correlation signal to the center of the galaxy distribution, with a more severe restriction when properly using the component separation process to recover the \hi signal.
% The difference between \hi + WN + Fg and GNILC-deb shows that the first method increases the statistical power, but more realistic analyzes are needed on parameter estimation.

%The influence of foreground contaminants of the 21\,cm signal on auto-APS is very strong for $\ell \gtrsim 210$, but weak in cross-APS until $\ell \lesssim 300$. 

%To evaluate the $b^2$ factor from the GNILC-deb scenario in more detail, we show in Figure \ref{FigBiasFg_Cls} its amplitude as a function of the multipole for the auto- and cross-APS of the BINGO $z$-bins number 14, 23, and 30, the closest to the center of the LSST galaxy distribution (Tab. \ref{Tab_Zbins}).
% illustrates the middle and lower rows of Figure \ref{FigBiasNoises}, i.e., the $b^2$ to auto- and cross-APS for input \hi maps and GNILC-deb maps.

\begin{figure}[!ht]
    \centering
    %\begin{tabular}{cc}
    %    \includegraphics[width=0.47\linewidth]{Figuras/BiasFg_Cls_14.pdf} &  \includegraphics[width=0.47\linewidth]{Figuras/BiasFg_Cls_23.pdf} 
    %\end{tabular}
    \includegraphics[width=0.65\linewidth]{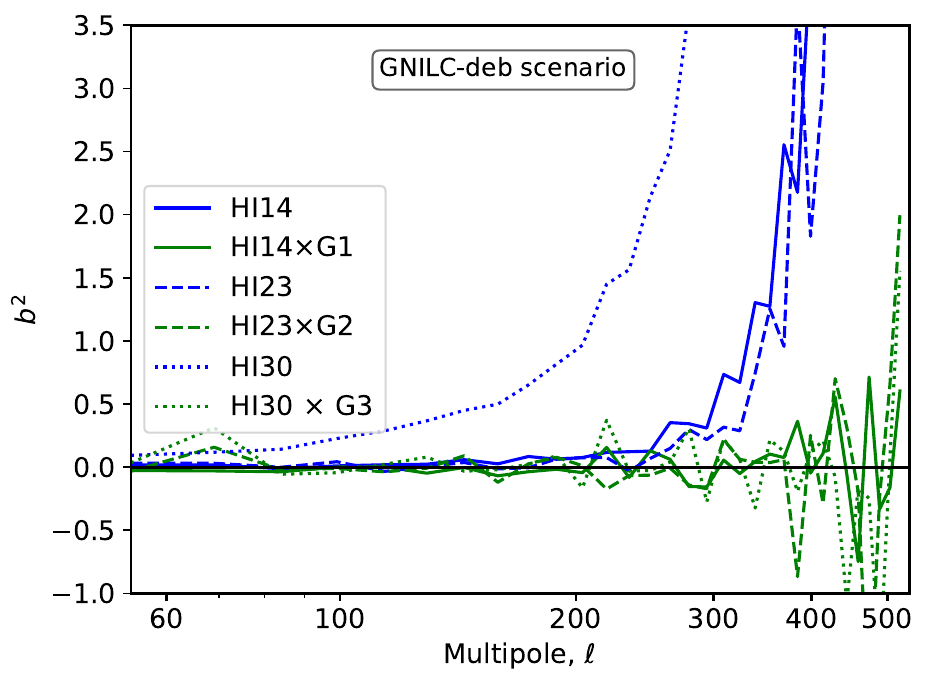}
    \caption{%$b^2$ factor for \hi auto-APS (dashed curves) and cross-APS (solid curves) using the \hi Input (blue curves) and GNILC-deb cases (green curves). The redshift bins are indicated in each panel (see Fig. \ref{Tab_Zbins}). Upper left panel: HI bin 14 and LSST bin 1 (as shown in Figure \ref{FigZbins}), $\Bar{z} \approx 0.25$.
    %\gabriel{
    $b^2$ factor for \hi auto-APS (blue curves) and cross-APS (green curves) for GNILC-deb scenario. %The solid, dashed, and dotted lines represent results using the three BINGO z-bins, 14, 23, and 30, respectively (see Tab. \ref{Tab_Zbins}).
    %\alex{
    The solid, dashed, and dotted lines represent results using the BINGO bins described in Table \ref{Tab_Zbins}.
    }
    %} 
    %}
    \label{FigBiasFg_Cls}
\end{figure}

Figure \ref{FigBiasFg_Cls} presents the $b^2$ amplitude as a function of the multipole for the auto- and cross-APS of the BINGO $z$-bins number 14, 23, and 30, the closest to the center of the LSST galaxy distribution (Tab. \ref{Tab_Zbins}).
This figure highlights how the auto-APS are significantly affected by residual contamination at multipoles $\ell \gtrsim 250-270$ (a scale-dependent distortion), with a more pronounced effect from the last redshift bin (the border effect associated with the GNILC method as mentioned previously).
On the other hand, the cross-APS amplitudes are considerably less affected by the residual contamination, exhibiting an approximately scale-independent $b^2$ factor. 
This enforces the power of the cross-correlation analyses; unlike the autocorrelation, where contamination dominates from relatively low multipoles, the cross-spectra, using \hi IM $z$-bins closest to the center of the photometric galaxy distribution, are much less affected at higher multipoles.
A similar result is observed by \cite{shi2020hir4}, where the authors also found a scale dependence of the $b^2$ factor with multipole for autocorrelation, and mild $\ell$-dependence for the cross-correlation. 

% We observe that, in the case of autocorrelation, residual foregrounds introduce a scale-dependent distortion, particularly at intermediate and high multipoles. In contrast, the bias in the cross-power spectra is closer to unity and displays only mild $\ell$-dependence, supporting the idea that uncorrelated residuals primarily affect the autocorrelation signal as observed by \cite{shi2020hir4}. %These results reinforce the robustness of the cross-correlation technique as a reliable method to recover cosmological information from 21\,cm intensity mapping observations, even in the presence of imperfect foreground removal.
%
\subsubsection{Statistical Significance}
\label{SecStatsSign}

\begin{figure}[!ht]
    \centering
    \includegraphics[width=\linewidth]{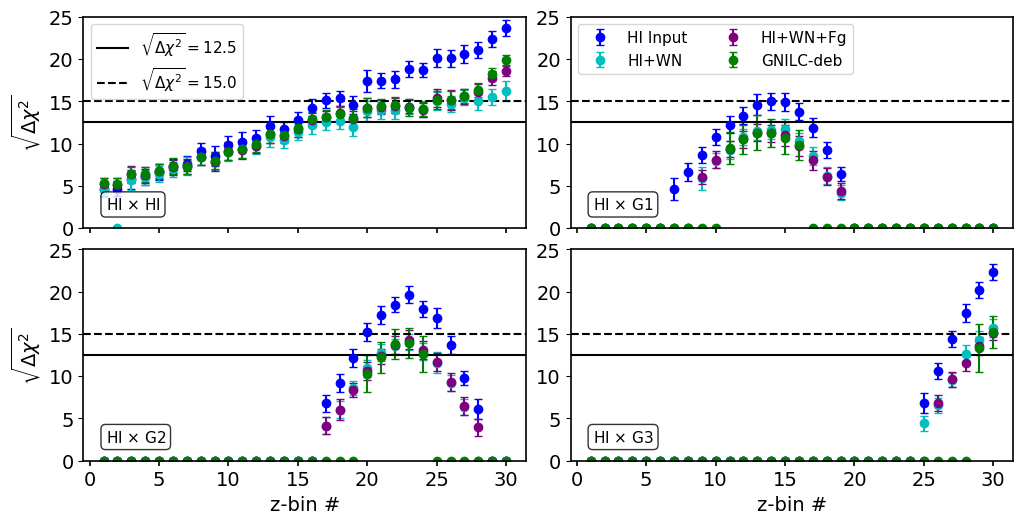}
    \caption{Statistical significance of the \hi signal detection through the autocorrelation of the 30 BINGO $z$-bins and their cross-correlation with each of the 3 LSST photo-$z$ bins (see Figure \ref{FigZbins}).  
    Upper left panel: auto-APS for the 30 BINGO z-bins.
    Upper right: cross-APS among all BINGO z-bins and LSST bin 1 ($\Bar{z}_G \approx 0.25$).
    Lower left: cross-APS among all BINGO z-bins and LSST bin 2 ($\Bar{z}_G \approx 0.35$).
    Lower right: cross-APS among all BINGO z-bins and LSST bin 3 ($\Bar{z}_G \approx 0.45$).
    In all the cases, $\sqrt{\Delta \chi^2}$ is calculated employing the multipole range $99 < \ell < 309$. 
    The solid and dashed horizontal lines denote $\sqrt{\Delta \chi^2} = 12.5$ and $15.0$, respectively, for reference.}
    \label{FigSignificancia}
\end{figure}

To quantify the statistical 
% assess the 
significance with which the APS would be detected, we compute $\sqrt{\Delta \chi^2} = \sqrt{\chi^2 -\chi^2_{\text{null}}}$, the difference between the $\chi^2$ value calculated for a given simulation using our model (Sec. \ref{SubsecAPSmodeling}), 
% distinct models or data configurations 
and using a null model with zero (auto- or cross-) APS. 
%, providing a quantitative measure of the improvement in the fit, in the same fashion as done by \cite{cunnington2023h}. 
For each scenario, the $\sqrt{\Delta \chi^2}$ value is estimated for 50 realizations, but using all the 1500 realizations to obtain the covariance matrix. %%% Muito bem lembrado, não tinha citado isso aqui. 
For the GNILC-deb realizations, we use the covariance matrix calculated from the 1500 fast simulations (\hi+WN+Fg). 
In all cases, we employ the multipole range $99 < \ell < 309$ (vertical dashed white lines in Fig. \ref{FigBiasNoises}), chosen to avoid multipole with poor SNR and to maximize the statistical significance. Figure \ref{FigSignificancia} shows the average and 68\% error bars from the $\sqrt{\Delta \chi^2}$ estimated for the autocorrelation of each of the $30$ BINGO $z$-bins and their cross-correlation with the three LSST photometric redshift bins, for the scenarios (i) to (iv) listed in Section \ref{SubsecAPS}. 

The statistical significance for the autocorrelation of each BINGO $z$-bin is shown in the first panel of Fig. \ref{FigSignificancia}, showing a larger value of $\sqrt{\Delta \chi^2}$ the higher the redshift. 
This result also shows that the detectability of the \hi signal at high redshifts is impacted by the inclusion of contaminants ($z \gtrsim 0.25$). 
The similar statistical significance among the three scenarios  with contaminants (ii-iv) indicates the thermal noise, present in all of them, as the main responsible for such degradation.
Notice that, within the BINGO redshift range, higher redshifts are more affected by white noise because it starts to dominate the cosmological signal at smaller multipoles compared to lower redshifts, as discussed by \cite{novaes2022bao}.

The remaining panels of Fig. \ref{FigSignificancia} show the statistical significance derived from the cross-correlation between of the 30 BINGO bins and each of the three LSST bins (cf. Figure \ref{FigZbins}). 
For each LSST bin, the peak of the significance occurs at the center of the galaxy photo-$z$ bins, where the number of galaxies contributing to both tracers is maximized and the cross-APS is more accurately estimated (see Section \ref{SubsecAPS} and Fig. \ref{FigBiasNoises}). 
Again, the three three scenarios with contaminants (ii-iv, the cyan, purple, and green dots) show similar statistical significance, indicating that the impact of foregrounds is negligible compared to the thermal noise, even with the application of additive debiasing correction. 
Despite the GNILC-deb maps (green dots) and the fast simulations (purple dots) yielding similar significance values around the center of the LSST photo-$z$ bins, the former is more restricted to center of the galaxy distribution, with some of the green dots further away of the center dropping to a significance close to zero.
This suggests that our fast simulations are not representative enough of the foreground cleaned maps, indicating that the degradation of the \hi signal by the component separation process impacts the cross-correlation when using \hi redshift bins away from the center of the galaxy distribution. 

Comparing the results shown in Fig. \ref{FigSignificancia} from auto- and cross-APS for \hi $z$-bins closer to the center of the photo-$z$ galaxy distributions, we find very similar statistical significance among them. 
This indicates that, although the cross-correlation signal, on average, is not degraded in amplitude due to the poor photometric redshift determination.  Differently from the results presented by \cite{cunnington2019impact}, the noise-like effect introduced by the presence of galaxies not contributing to the \hi $z$-bin, as well as the presence of residual contamination in the \hi maps, results in a significant deterioration  of the statistical significance of their detection. 
This deterioration is sufficient to make the detection of the cross-correlation signal as likely as the autocorrelation, even with the great advantage of the systematics being uncorrelated to the cosmological signal. 

However, it is important to remember that, as a first analysis addressing the use of the full photometric galaxy distribution for cross-correlation with \hi IM, our work employ semi-realistic simulations of the future BINGO observations. 
As such, they lack to account for instrumental effects relevant for intensity mapping experiments, such as the low frequency ($1/f$) noise \citep{2018/harper} and the leakage of polarized foregrounds into the intensity data \citep{alonso2015blind, 2021/cunnington}. 
Additional sources of contamination will certainly impact the \hi autocorrelation signal more, 
but should have a weaker impact in the  cross-correlation signal.
Additionally, increasing the number of frequency bands should improve the foreground cleaning process \citep{mericia2023testing}, providing better auto- and cross-APS estimates, particularly at higher redshifts. 
This would likely allow improving the statistical significance for the cross-correlation detection using \hi $z$-bins further away from the center of the galaxy distributions.

\section{Parameter estimation}\label{SecParams}

As demonstrated by several authors, the cross-correlation of \hi IM observations with galaxy catalogs can provide an estimate of the cosmological \hi signal \citep{masui2013measurement, wolz2016intensity, wolz2017determining, wolz2019intensity, Wolz_gbt:2022, Jiang:2023, cunnington2023h, amiri2023detection, mazumder2025hi, zheng2025cosmology}. 
% In this section, we assess the potential of the combination of BINGO and LSST surveys in estimating the product among the degenerate astrophysical parameters, $b_{\text{\hi}}$, $\Omega_\text{\hi}$, and the cross-correlation coefficient, $r$,
% % In this section, we describe the procedure used to estimate the astrophysical parameters 
% that characterize the 21\,cm signal. 
In this section, we assess the potential of  combining simulations of BINGO and LSST data to estimate the product of the degenerate parameters $b_{\text{\hi}}$, $\Omega_\text{\hi}$, and the cross-correlation coefficient, $r$,
% In this section, we describe the procedure used to estimate the astrophysical parameters 
that characterize the 21\,cm signal. 

The aforementioned papers, as well as other works in the literature reporting constraints on $b_{\text{\hi}} \Omega_\text{\hi} r$, %as far as we know, 
model the matter clustering through the power spectrum in Fourier space. 
In this work, we show that the APS is a promising alternative to be used in the same task. 
Following \cite{cunnington2023h}, we assume %both parameters, 
$b_{\text{\hi}}$ and $\Omega_\text{\hi}$ to be approximately constant inside the redshift bins, given the narrow width of the $z$-bin; simulations are generated based on this assumption. 
In this way, one can move this quantity outside the integrals in Eq. \ref{eq:clls_theo}, such that we have, respectively, for autocorrelation and cross-correlation, 
% Matter clustering has been measured through \hi IM mainly using cross-correlations, constraining the degenerate astrophysical parameter $\Omega_{\hi}b_{\hi}r$ by measuring the right-hand side of Equation \ref{eq:clls_theo} and dividing it by the modeled matter power spectrum with a fixed galaxy bias (Table \ref{Tab_Zbins}). 
% Nonetheless, such measurements can suffer from wide-angle biasing when transform the observed set of pixalised tomographic maps to a cartesian grid, enabling the use of FFT\footnote{See \cite{cunnington2023h} for corrections} so the projection kernel remains the same for the matter fluctuations on the sphere, allowing to properly compare with the cross-APS:
%
\begin{align}
    C_{\text{\hi},\, \ell}^{ij}\, & = (\Omega_{\text{\hi}} \, b_{\text{\hi}})^2 \,\, C_{\rm m,\, \ell}^{ij} \label{eq:cl_model_auto} \\ 
    C_{\text{\hi} \, g,\, \ell}^{ij}\,& = \Omega_{\text{\hi}} \, b_{\text{\hi}} \, r_\ell \,\, C_{\rm \times g,\, \ell}^{ij},
    \label{eq:cl_model_cross} 
\end{align}
%
% where $\mathcal{A}_{\ell,X}^{ij}$ is the angular cross-correlation amplitude (ACCA) and 
where the degenerate quantity, $b_{\text{\hi}} \Omega_{\text{\hi}} r$, is assumed to be scale independent, affecting only the amplitude of the cross-APS model.
It is worth noting that the correlation coefficient $r$ that  accounts for the stochasticity between the two cosmological fields should be scale dependent \citep{cunnington2023h, Wolz_gbt:2022, wolz2016intensity, masui2013measurement}. 
However, since it approaches unity on large scales, the scale independency is a reasonable assumption on the scales considered here, $r_\ell = r$. 
%This way, 
In this case, $b_{\text{\hi}} \, \Omega_{\text{\hi}} \, r$ and $b_{\text{\hi}} \, \Omega_{\text{\hi}}$ can be estimated by fitting, respectively, the amplitude of the auto-APS ($C_{\text{\hi},\, \ell}^{ij}$,)  and cross-APS ($C_{\text{\hi} \, g,\, \ell}^{ij}$) and  obtained  from the simulations. 

For autocorrelation, the model $C_{\rm m,\, \ell}^{ij}$ corresponds to the matter APS, calculated using the {\tt UCLC$\ell$} code and imposing $b_{\text{\hi}} = \Omega_{\text{\hi}} = 1$. 
For cross-correlation, the model $C_{\rm \times g,\, \ell}^{ij}$ corresponds to the cross-APS and already accounts for the galaxy bias $b_G$ and the redshift distribution $n(z)$, both considered as known, given that they can be obtained from  galaxy surveys (see, e.g., \cite{zhang2022transitioning}), again imposing $b_{\text{\hi}} = \Omega_{\text{\hi}} = 1$.

%To fit the cross-APS, $C_{\rm \times g,\, \ell}^{ij}$, already accounts for the galaxy bias, $b_G$ and the redshift distribution, $n(z)$, both considered as known, given that they can be obtained from  galaxy surveys (see, e.g., \cite{zhang2022transitioning}).

%Specifically, w
We constrain the degenerate quantities, $b_{\text{\hi}} \, \Omega_{\text{\hi}} \, r$ and $b_{\text{\hi}} \, \Omega_{\text{\hi}}$, through a least-square fit to the amplitude of the auto- and cross-APS obtained from the simulations, similarly to what is done by \cite{cunnington2023h}.
We minimize\footnote{Implemented using the {\tt curve\_fit} function from {\tt SciPy} \citep{virtanen2020scipy}.} the $\chi^2$ defined as
\begin{equation} 
    \chi^2(\lambda) = \sum_{k, m} \left[ X^{i}_{k} - \lambda x^{i}_{k} \right] \left[ C^{ij}_{km} \right]^{-1} \left[ X^{j}_{m} - \lambda x^{j}_{m} \right], 
\end{equation}
where $\lambda$ is the degenerate quantity to be estimated, $i,j$ refers to the redshift bins, and $k,m$ indices run over the different multipoles bands, $\ell$. 
% of the power spectra $C_\ell$, 
Following our findings from last section, 
% Based on the SNR $> 1$ range (Figure \ref{FigSNRClsComparingAutoCros}) 
the fitting is performed using the multipole interval $99 \lesssim \ell \lesssim 309$, which is the less affected by residual contamination (best SNR).
The terms $X^{i}_{k}$ and $x^{i}_{k}$ correspond to the measured APS, $C_{\text{\hi},\, \ell}^{ij}$ or $C_{\text{\hi} \, g,\, \ell}^{ij}$, and model, $C_{\rm m,\, \ell}^{ij}$ or $C_{\rm \times g,\, \ell}^{ij}$, respectively;
% , with the latter computed using the {\tt UCLC$\ell$} code.
$\left[ C^{ij}_{km} \right]^{-1}$, corresponds to the rescaled precision matrix\footnote{To mitigate bias in the inverse covariance matrix due to statistical noise, we adopt a rescaled precision matrix as proposed in \cite{hartlap2007why}. Given the number of simulations, $N$, and the number of entries of the APS data vectors, $N_p$, the rescaled precision matrix is obtained by $[C_{ij}^{km}]^{-1} \rightarrow (N - N_p -2)/(N - 1) \, [C_{ij}^{km}]^{-1}$.}. 

\begin{figure*}[t]
    \centering
    \includegraphics[width=\linewidth]{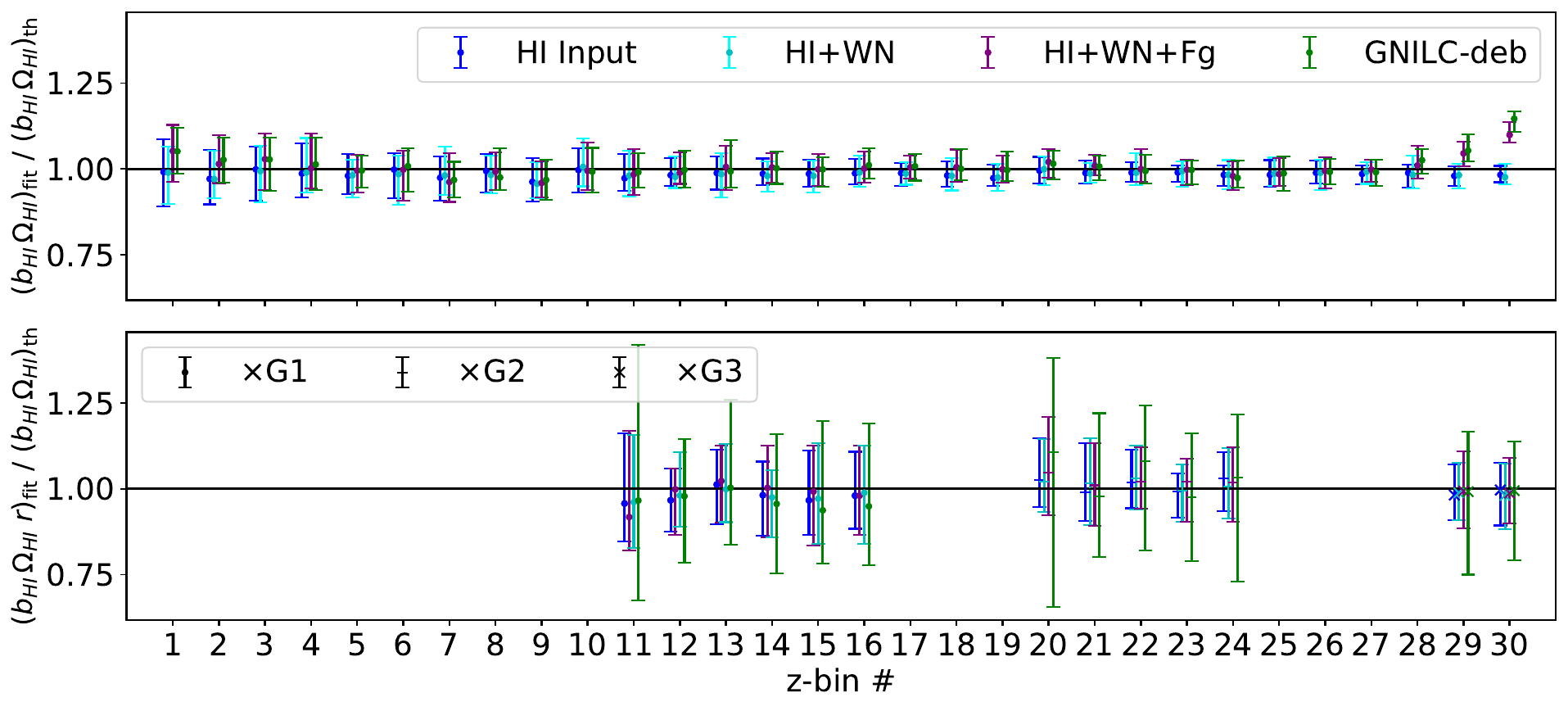}
    \caption{Median and 16 and 84 percentiles (1$\sigma$ dispersion) of the $b_{\text{\hi}} \, \Omega_{\text{\hi}}$ and $b_{\text{\hi}} \, \Omega_{\text{\hi}} \, r$ quantities constrained, respectively, from  
    % Parameter recovery from 
    auto-APS (top panel) and cross-APS (bottom panel) measured from simulations of scenarios (i) to (vi). The x-axis represents the BINGO redshift bin, and the y-axis shows the ratio between the estimated and theoretical values. 
    The bottom panel shows only the results from cross-APS whose statistical significance is higher than unity (green dots in Figure \ref{FigSignificancia}). Results from the cross-correlation with the first, second, and third LSST bins are represented by $\bullet$, $+$, and $\times$ symbols, respectively. Results are slightly shifted to right for better visualization.  
    % The cross-correlation dots are selected from significance detection of each LSST redshift bin (Figure \ref{FigSignificancia}): $\bullet$ for the first, $+$ for the second and $\times$ for the last.
    }
    \label{FigParams}
\end{figure*}

Figure \ref{FigParams} compares the estimated values of the degenerate quantities, from fitting both auto- and cross-APS amplitudes, for 50 realizations from each of the four scenarios. 
For the cross-correlation, we show only the results obtained for the BINGO $z$-bins closest to the center of the LSST galaxy distribution, that is, for those cross-APS leading to statistical significance higher than unity in Fig. \ref{FigSignificancia}.

The results of the present analysis suggest that the application of cross-correlation using our current semi-realistic simulations, the GNILC-deb realizations, introduces no observable systematic error in the average $b_{\text{\hi}} \Omega_{\text{\hi}} r$ values when benchmarked against the \hi input scenario. 
In contrast, the autocorrelation analyses indicates a displacement in the average $b_{\text{\hi}} \Omega_{\text{\hi}}$ toward larger values at the higher redshifts, a deviation plausibly caused by enhanced the amplitude of the foreground residuals within the corresponding redshift intervals.

Still in Fig. \ref{FigParams}, we see that, although the cross-correlation results do not show any shift, it leads to a larger 1$\sigma$ dispersion of the estimated values compared to thje autocorrelation results. 
Considering the \hi Input scenario (blue bars), the cross-correlation with the first, second, and third photo-$z$ bins lead, respectively, to 1$\sigma$ dispersion at least $2.07$, $2.15$ and $2.58$ times larger than those from the autocorrelation of the corresponding BINGO $z$-bins. 
For the GNILC-deb scenario (green bars), the 1$\sigma$ dispersion from cross-correlation increases by $2.83$, $3.44$ and $3.69$, respectively, with respect to autocorrelation. 
Such findings from the GNILC-deb case show the combined effect of the photo-$z$ error and the deterioration of the \hi signal by the component separation process. 
Additionally, we show that \hi+WN, \hi+WN+Fg and GNILC-deb scenarios do not seem to impact significantly the \hi Input fitting results for autocorrelation.

Finally, the GNILC-deb scenario leads to 1$\sigma$ dispersion from $1.16$ to $3.65$ times larger than those from \hi+WN+Fg for cross-correlation. 
For the BINGO $z$-bins closest to the center of LSST distributions (see Tab. \ref{Tab_Zbins}), the 1$\sigma$ uncertainty in the degenerate quantities, estimated from the GNILC-deb APS, correspond to a fraction of 0.05 and 0.20, respectively, of the theoretical $\,b_{\text{\hi}} \Omega_{\text{\hi}}$ and $b_{\text{\hi}} \Omega_{\text{\hi}} r$ values, from auto and cross-correlation. 
Since in all the cases the 1$\sigma$ intervals (68\% of the simulations) are far from zero, as can be seen in Fig. \ref{FigParams}, our results suggest that it would be possible to detect the \hi signal and constrain the degenerate parameters, considering the semi-realistic simulations employed here. 
Nevertheless, we state that our conclusions might be challenged by real \hi data or more complete simulations, conditions under which we expect the autocorrelation results to be more impacted, as discussed in Sec. \ref{SecStatsSign}. 

\section{Summary and conclusions}\label{SecConclusions}

This work assesses, for the first time, the detectability of the 21\,cm signal as measured by the BINGO radiotelescope, in its Phase 1 of operation, through cross-correlation with galaxy catalogs. 
In particular, we explore the feasibility of using photometric surveys for this purpose, considering specifically here the future LSST survey, which presents an excellent overlap in area and redshift with BINGO. 
Cross-correlation between \hi IM and photometric surveys has not been deeply investigated due to poor photo-$z$ constraints, with a clear preference in the literature for spectroscopic measurements. 
However, given the large number of current, high-quality, photometric surveys available, we believe that their synergy with \hi IM surveys should be better evaluated. 
Here, we explicitly show how photo-$z$ errors would impact the cross-correlation measurements, focusing on employing the full photometric galaxy distribution (Fig. \ref{FigZbins}).

This is a novel approach, where we cross-correlate each of the 30 BINGO $z$-bins with the three LSST photo-$z$ bins, in a complementary fashion of what is presented, e.g., in \cite{cunnington2019impact}, where the photometric galaxies are selected to match the \hi bin width. 
Our analyses employ the APS as a clustering estimator, suitable in the case of pixelized line intensity maps, and rely on BINGO-like and LSST-like simulations. 
We analyze the sky region delimited by the BINGO footprint (fully covered by LSST), investigating also the impact of including thermal noise and foreground residuals. In contrast with results from \cite{cunnington2019impact}, we do not find a significant degradation of the cross-correlation signal, on average, likely because considering the full photometric redshift bin increases the number of galaxies also contributing to the \hi signal. 

However, we observe that the 68\% region from the APS calculated from simulations is wider for cross-correlation compared to autocorrelation, which we suspect is due to the photo-$z$ error. 
To investigate this point, we implemented a null test, in which we cross-correlate a \hi map with a set of Gaussian random maps, mimicking the presence of galaxies in the photometric sample not contributing to the \hi signal.
This test indicates that such galaxies introduce a noise-like behavior in the cross-APS, explaining why the auto-APS shows a higher SNR. Additionally, it is interesting to notice from our results that such behavior is similar to the impact from adding thermal noise and foregrounds residuals, which also boost the noise in auto and cross-correlation. 

However, since the \hi contaminants do not correlate with the galaxies, no bias is observed in the average amplitude of the cross-APS. 
On the other hand, for \hi autocorrelation,  although the 68\% regions are smaller, a clear bias is found at larger multipoles, whose amplitude and affected multipoles vary with the redshift bin. 
Based on our study of the SNR from both auto and cross-APS, we selected the highest SNR multipole range to assess the statistical significance of the APS detection. 
We find that, given the noise-like behavior introduced by the photo-$z$ error, the statistical significance of the cross-correlation from \hi bins closest to the center of the photo-$z$ distribution is reduced to levels comparable to that from autocorrelation. 

In fact, the significance gets smaller the further away is the $z$-bin from the center of the galaxy distribution.
This is expected, given that the number of galaxies contributing to both tracers gets smaller the farther the \hi bin is from the center of the photo-$z$ distribution, decreasing the amplitude of the expected cross-APS and increasing the noise at the same time.

Finally, considering the same multipole range, we also evaluate how well actual BINGO observations would provide constraints on the degenerate quantities $b_{\text{\hi}} \Omega_{\text{\hi}}$ and $b_{\text{\hi}} \Omega_{\text{\hi}} r$ from auto and cross-correlation, respectively. 
We find that the autocorrelation should bias the estimates towards higher redshifts, where the foreground residual is more relevant. 
From cross-correlation, considering only the cross-APS that could be detected according to our statistical significance analysis, no bias is observed. 
However, the constrained values are impacted by the photo-$z$ error and the presence of contaminants, showing a noise-like behavior, with a larger 1$\sigma$ dispersion of the estimated values compared to results from autocorrelation. 
Still, both auto and cross-correlation results indicate that constraints would be possible, considering the semi realistic simulations considered here.

It is worth emphasizing that, while our analysis brings new findings to the context of using cross-correlation to detect the \hi signal and the use of photometric surveys for this task, the synergy between BINGO and LSST deserves further study. 
Accounting for effects from  beam shape, $1/f$ noise, and polarization leakage, should have a more significant impact on autocorrelation. 
Also, a thinner slicing of the BINGO frequency band, that is, a larger number of $z$-bins, would also contribute to a better performance of the foreground cleaning process, allowing to improve auto- and cross-APS estimates. 
Therefore, our results suggest that the refinement of the current analyses, in the light of real data, should produce very interesting results.

Another interesting aspect we plan to investigate in future analyses is the possibility of improving the SNR from cross-correlations by combining the \hi $z$-bins encompassed by the photometric galaxy distribution. 
Summing all the radio redshift bins inside a given photo-z interval might increase the likelihood of detecting the \hi signal through cross-correlation with photometric surveys, presenting this method as an additional tool to be used for cross-correlation analyses.

Finally, we stress that even using BINGO-like and LSST-like simulations, which is not the most realistic scenario, our results indicate that the combination of current and future photometric and radio IM data have quite a good potential to detect the \hi signal through cross-correlation. 
This motivates future investigation towards a more complete framework to assess the full capability of cross-correlating photometric galaxy surveys with \hi IM experiments, BINGO and LSST being our best bet for the near future.

% \appendix
% \section{Some title}
% Please always give a title also for appendices.

\acknowledgments

%\alex{
G.A.S.S. acknowledges resources received from CAPES under process 88887.602823/2021-00. 
C.P.N. thanks Serrapilheira Institute and São Paulo Research Foundation (FAPESP; grant 2019/06040-0) for financial support.
C.A. Wuensche thanks CNPq for grants 407446/2021-4 and 312505/2022-1, the Brazilian Ministry of Science, Technology and Innovation (MCTI) and the Brazilian Space Agency (AEB) who supported the present work under the PO 20VB.0009.
B.B.B. acknowledges resources received from FAPESP under processes 2021/04786-5, 2022/16749-0 and 2024/12902-3. 
E.J.M. thanks INPE for the financial support provided by the PCI fellowship. The authors thank V. Liccardo and R. Mokeddem for useful discussions and  manuscript reading. The BINGO project is supported by FAPESQ-PB, the State of Paraiba, FINEP, and FAPESP-SP, Brazil, by means of several grants. The results of this work used  the software packages {\tt astropy} \citep{whelan2018astropy}, {\tt healpy} \citep{zonca2019healpy}, {\tt numpy} \citep{2011/numpy}, {\tt scipy} \citep{virtanen2020scipy}, {\tt matplotlib} \citep{2007/matplotlib}, {\tt UCLC$\ell$} \citep{loureiro2019cosmological} and {\tt FLASK} \citep{xavier2016flask}.
%}
%\paragraph{Note added.} This is also a good position for notes added
%after the paper has been written.

% The bibliography will probably be heavily edited during typesetting.
% We'll parse it and, using the arxiv number or the journal data, will
% query inspire, trying to verify the data (this will probalby spot
% eventual typos) and retrive the document DOI and eventual errata.
% We however suggest to always provide author, title and journal data:
% in short all the informations that clearly identify a document.

\bibliography{refs} % your references Yourfile.bib

\providecommand{\href}[2]{#2}\begingroup\raggedright\begin{thebibliography}{10}

\bibitem{2022:JamesWebb-Robertson}
B.E.~{Robertson}, \emph{{Galaxy Formation and Reionization: Key Unknowns and
  Expected Breakthroughs by the James Webb Space Telescope}},
  \href{https://doi.org/10.1146/annurev-astro-120221-044656}{\emph{Ann.\ Rev.\
  Astron.\ Astrophys.} {\bfseries 60} (2022) 121}
  [\href{https://arxiv.org/abs/2110.13160}{{\ttfamily 2110.13160}}].

\bibitem{amendola2018cosmology}
L.~Amendola, S.~Appleby, A.~Avgoustidis, D.~Bacon, T.~Baker, M.~Baldi et~al.,
  \emph{Cosmology and fundamental physics with the {E}uclid satellite},
  {\emph{Living rev.\ relativ.} {\bfseries 21} (2018) 1}.

\bibitem{ivezic2019lsst}
{\v{Z}}.~Ivezi{\'c}, S.M.~Kahn, J.A.~Tyson, B.~Abel, E.~Acosta, R.~Allsman
  et~al., \emph{{LSST}: from science drivers to reference design and
  anticipated data products}, {\emph{The Astrophysical Journal} {\bfseries 873}
  (2019) 111}.

\bibitem{chen2012tianlai}
X.~Chen, \emph{The tianlai project: a 21cm cosmology experiment},  in
  \emph{Int.\ J.\ Mod.\ Phys.\ Conf.\ Ser.}, vol.~12, pp.~256--263, World
  Scientific, 2012.

\bibitem{2022/hirax}
D.~Crichton, M.~Aich, A.~Amara, K.~Bandura, B.A.~Bassett, C.~Bengaly et~al.,
  \emph{Hydrogen intensity and real-time analysis experiment: 256-element array
  status and overview}, {\emph{J. Astron. Telesc. Instrum. Syst.} {\bfseries 8}
  (2022) 011019}.

\bibitem{2011/fast}
R.~Nan, D.~Li, C.~Jin, Q.~Wang, L.~Zhu, W.~Zhu et~al., \emph{The
  five-hundred-meter aperture spherical radio telescope {(FAST)} project},
  {\emph{Int.\ J.\ Mod.\ Phys.\ D} {\bfseries 20} (2011) 989}.

\bibitem{2020/ska}
D.J.~Bacon, R.A.~Battye, P.~Bull, S.~Camera, P.G.~Ferreira, I.~Harrison et~al.,
  \emph{Cosmology with phase 1 of the square kilometre array red book 2018:
  technical specifications and performance forecasts}, {\emph{Publications of
  the Astronomical Society of Australia} {\bfseries 37} (2020) e007}.

\bibitem{battye2013h}
R.~Battye, I.~Browne, C.~Dickinson, G.~Heron, B.~Maffei and A.~Pourtsidou,
  \emph{{HI} intensity mapping: a single dish approach}, {\emph{Mon.\ Not.\ R.\
  Astron.\ Soc.} {\bfseries 434} (2013) 1239}.

\bibitem{2015/bigot-sazy}
M.-A.~Bigot-Sazy, C.~Dickinson, R.A.~Battye, I.~Browne, Y.-Z.~Ma, B.~Maffei
  et~al., \emph{Simulations for single-dish intensity mapping experiments},
  {\emph{Mon.\ Not.\ R.\ Astron.\ Soc.} {\bfseries 454} (2015) 3240}.

\bibitem{abdalla2022bingoI}
E.~Abdalla, E.G.~Ferreira, R.G.~Landim, A.A.~Costa, K.S.~Fornazier,
  F.B.~Abdalla et~al., \emph{The {BINGO project-I}. baryon acoustic
  oscillations from integrated neutral gas observations}, {\emph{Astron.\
  Astrophys.} {\bfseries 664} (2022) A14}.

\bibitem{2017/santos}
M.G.~Santos, M.~Cluver, M.~Hilton, M.~Jarvis, G.I.~Jozsa, L.~Leeuw et~al.,
  \emph{Meerklass: Meerkat large area synoptic survey}, {\emph{arXiv preprint
  arXiv:1709.06099} (2017) }.

\bibitem{2014/chime}
K.~Bandura, G.E.~Addison, M.~Amiri, J.R.~Bond, D.~Campbell-Wilson, L.~Connor
  et~al., \emph{Canadian hydrogen intensity mapping experiment (chime)
  pathfinder},  in \emph{Ground-based and Airborne Telescopes V}, vol.~9145,
  p.~914522, SPIE, 2014.

\bibitem{Tegmark:2003}
M.~Tegmark, A.~de~Oliveira-Costa and A.J.S.~Hamilton, \emph{High resolution
  foreground cleaned cmb map from wmap},
  \href{https://doi.org/10.1103/PhysRevD.68.123523}{\emph{Phys. Rev. D}
  {\bfseries 68} (2003) 123523}.

\bibitem{alonso2015blind}
D.~Alonso, P.~Bull, P.G.~Ferreira and M.G.~Santos, \emph{Blind foreground
  subtraction for intensity mapping experiments}, {\emph{Mon.\ Not.\ R.\
  Astron.\ Soc.} {\bfseries 447} (2015) 400}.

\bibitem{olivari2016extracting}
L.~Olivari, M.~Remazeilles and C.~Dickinson, \emph{Extracting {H{\sc i}}
  cosmological signal with generalized needlet internal linear combination},
  {\emph{Mon.\ Not.\ R.\ Astron.\ Soc.} {\bfseries 456} (2016) 2749}.

\bibitem{2020/carucci}
I.P.~Carucci, M.O.~Irfan and J.~Bobin, \emph{Recovery of 21-cm intensity maps
  with sparse component separation}, {\emph{Mon.\ Not.\ R.\ Astron.\ Soc.}
  {\bfseries 499} (2020) 304}.

\bibitem{wuensche2022bingo}
C.A.~Wuensche, T.~Villela, E.~Abdalla, V.~Liccardo, F.~Vieira, I.~Browne
  et~al., \emph{The {BINGO project-II}. instrument description},
  {\emph{Astron.\ Astrophys.} {\bfseries 664} (2022) A15}.

\bibitem{2018/harper}
S.~Harper, C.~Dickinson, R.~Battye, S.~Roychowdhury, I.~Browne, Y.-Z.~Ma
  et~al., \emph{Impact of simulated 1/f noise for hi intensity mapping
  experiments}, {\emph{Mon.\ Not.\ R.\ Astron.\ Soc.} {\bfseries 478} (2018)
  2416}.

\bibitem{2021/li}
Y.~Li, M.G.~Santos, K.~Grainge, S.~Harper and J.~Wang, \emph{H i intensity
  mapping with meerkat: 1/f noise analysis}, {\emph{Mon.\ Not.\ R.\ Astron.\
  Soc.} {\bfseries 501} (2021) 4344}.

\bibitem{ding_beam:2024}
J.~Ding, X.~Wang, U.-L.~Pen and X.-D.~Li, \emph{Correlation-based beam
  calibration of 21 cm intensity mapping}, {\emph{Astrophys.\ J.\ Supp.}
  {\bfseries 274} (2024) 44}.

\bibitem{Wang_calibration:2021}
J.~Wang, M.G.~Santos, P.~Bull, K.~Grainge, S.~Cunnington, J.~Fonseca et~al.,
  \emph{{Hi intensity mapping with MeerKAT: calibration pipeline for multidish
  autocorrelation observations}},
  \href{https://doi.org/10.1093/mnras/stab1365}{\emph{Mon.\ Not.\ R.\ Astron.\
  Soc.} {\bfseries 505} (2021) 3698}.

\bibitem{Chang:2010}
T.-C.~{Chang}, U.-L.~{Pen}, K.~{Bandura} and J.B.~{Peterson}, \emph{An
  intensity map of hydrogen 21-cm emission at redshift z $\sim $ 0.8},
  \href{https://doi.org/10.1038/nature09187}{\emph{Nature} {\bfseries 466}
  (2010) 463}.

\bibitem{switzer2013determination}
E.~Switzer, K.~Masui, K.~Bandura, L.-M.~Calin, T.-C.~Chang, X.-L.~Chen et~al.,
  \emph{Determination of $z \sim$ 0.8 neutral hydrogen fluctuations using the
  21 cm intensity mapping autocorrelation}, {\emph{Mon.\ Not.\ R.\ Astron.\
  Soc.} {\bfseries 434} (2013) L46}.

\bibitem{masui2013measurement}
K.~Masui, E.~Switzer, N.~Banavar, K.~Bandura, C.~Blake, L.-M.~Calin et~al.,
  \emph{Measurement of 21 cm brightness fluctuations at $z \sim 0.8$ in
  cross-correlation}, {\emph{Astrophys.\ J.} {\bfseries 763} (2013) L20}.

\bibitem{Wolz_gbt:2022}
L.~Wolz, A.~Pourtsidou, K.W.~Masui, T.-C.~Chang, J.E.~Bautista, E.-M.~Müller
  et~al., \emph{{Hi constraints from the cross-correlation of eBOSS galaxies
  and Green Bank Telescope intensity maps}},
  \href{https://doi.org/10.1093/mnras/stab3621}{\emph{Mon.\ Not.\ R.\ Astron.\
  Soc.} {\bfseries 510} (2021) 3495}.

\bibitem{cunnington2023h}
S.~Cunnington, Y.~Li, M.G.~Santos, J.~Wang, I.P.~Carucci, M.O.~Irfan et~al.,
  \emph{{HI} intensity mapping with meerkat: power spectrum detection in
  cross-correlation with wigglez galaxies}, {\emph{Mon.\ Not.\ R.\ Astron.\
  Soc.} {\bfseries 518} (2023) 6262}.

\bibitem{amiri2023detection}
M.~Amiri, K.~Bandura, T.~Chen, M.~Deng, M.~Dobbs, M.~Fandino et~al.,
  \emph{Detection of cosmological 21 cm emission with the canadian hydrogen
  intensity mapping experiment}, {\emph{Astrophys.\ J.} {\bfseries 947} (2023)
  16}.

\bibitem{paul2023first}
S.~Paul, M.G.~Santos, Z.~Chen and L.~Wolz, \emph{A first detection of neutral
  hydrogen intensity mapping on mpc scales at $z \approx 0.32$ and $z \approx
  0.44$}, {\emph{arXiv preprint arXiv:2301.11943} (2023) }.

\bibitem{dodelson2003modern}
S.~Dodelson, \emph{Modern cosmology}, Academic Press, Massachusetts, USA
  (2003).

\bibitem{gao2023asymptotic}
Z.~Gao, A.~Raccanelli and Z.~Vlah, \emph{Asymptotic connection between full-and
  flat-sky angular correlators}, {\emph{Phys.\ Rev.\ D} {\bfseries 108} (2023)
  043503}.

\bibitem{carucci2024hydrogen}
I.P.~Carucci, J.L.~Bernal, S.~Cunnington, M.G.~Santos, J.~Wang, J.~Fonseca
  et~al., \emph{Hydrogen intensity mapping with meerkat: Preserving
  cosmological signal by optimising contaminant separation}, {\emph{arXiv
  preprint arXiv:2412.06750} (2024) }.

\bibitem{Blake:2019}
C.~Blake, \emph{{Power spectrum modelling of galaxy and radio intensity maps
  including observational effects}},
  \href{https://doi.org/10.1093/mnras/stz2145}{\emph{Mon.\ Not.\ R.\ Astron.\
  Soc.} {\bfseries 489} (2019) 153}.

\bibitem{cunnington2024accurate}
S.~Cunnington and L.~Wolz, \emph{Accurate fourier-space statistics for line
  intensity mapping: Cartesian grid sampling without aliased power},
  {\emph{Mon.\ Not.\ R.\ Astron.\ Soc.} {\bfseries 528} (2024) 5586}.

\bibitem{benabou2024wide}
J.N.~Benabou, I.~Sands, H.S.G.~Gebhardt, C.~Heinrich and O.~Dor{\'e},
  \emph{Wide-angle effects in the power spectrum multipoles in next-generation
  redshift surveys}, {\emph{Phys.\ Rev.\ D} {\bfseries 110} (2024) 083526}.

\bibitem{Pourtsidou:2015}
A.~Pourtsidou, D.~Bacon and R.~Crittenden, \emph{Cross-correlation cosmography
  with intensity mapping of the neutral hydrogen 21 cm emission},
  \href{https://doi.org/10.1103/PhysRevD.92.103506}{\emph{Phys. Rev. D}
  {\bfseries 92} (2015) 103506}.

\bibitem{2019/padmanabhan}
H.~Padmanabhan, A.~Refregier and A.~Amara, \emph{Impact of astrophysics on
  cosmology forecasts for 21 cm surveys}, {\emph{Mon.\ Not.\ R.\ Astron.\ Soc.}
  {\bfseries 485} (2019) 4060}.

\bibitem{padmanabhan2020cross}
H.~Padmanabhan, A.~Refregier and A.~Amara, \emph{Cross-correlating 21 cm and
  galaxy surveys: implications for cosmology and astrophysics}, {\emph{Mon.\
  Not.\ R.\ Astron.\ Soc.} {\bfseries 495} (2020) 3935}.

\bibitem{shi2020hir4}
F.~Shi, Y.-S.~Song, J.~Asorey, D.~Parkinson, K.~Ahn, J.~Yao et~al., \emph{Hir4:
  cosmological signatures imprinted on the cross-correlation between a 21-cm
  map and galaxy clustering}, {\emph{Mon.\ Not.\ R.\ Astron.\ Soc.} {\bfseries
  499} (2020) 4613}.

\bibitem{cunnington2019impact}
S.~Cunnington, L.~Wolz, A.~Pourtsidou and D.~Bacon, \emph{Impact of foregrounds
  on {HI} intensity mapping cross-correlations with optical surveys},
  {\emph{Mon.\ Not.\ R.\ Astron.\ Soc.} {\bfseries 488} (2019) 5452}.

\bibitem{2019/abbott-des}
T.~Abbott, F.~Abdalla, A.~Alarcon, S.~Allam, F.~Andrade-Oliveira, J.~Annis
  et~al., \emph{Dark energy survey year 1 results: Measurement of the baryon
  acoustic oscillation scale in the distribution of galaxies to redshift 1},
  {\emph{Mon.\ Not.\ R.\ Astron.\ Soc.} {\bfseries 483} (2019) 4866}.

\bibitem{benitez2014j}
N.~Benitez, R.~Dupke, M.~Moles, L.~Sodre, J.~Cenarro, A.~Marin-Franch et~al.,
  \emph{J-pas: the javalambre-physics of the accelerated universe astrophysical
  survey}, {\emph{arXiv preprint arXiv:1403.5237} (2014) }.

\bibitem{hikage2019cosmology}
C.~Hikage, M.~Oguri, T.~Hamana, S.~More, R.~Mandelbaum, M.~Takada et~al.,
  \emph{Cosmology from cosmic shear power spectra with subaru hyper suprime-cam
  first-year data}, {\emph{Publ.\ of the Astr.\ Soc.\ Japan} {\bfseries 71}
  (2019) 43}.

\bibitem{serrano2023physics}
S.~Serrano, E.~Gazta{\~n}aga, F.J.~Castander, M.~Eriksen, R.~Casas,
  D.~Navarro-Giron{\'e}s et~al., \emph{The physics of the accelerating universe
  survey: narrow-band image photometry}, {\emph{Monthly Notices of the Royal
  Astronomical Society} {\bfseries 523} (2023) 3287}.

\bibitem{loureiro2019cosmological}
A.~Loureiro, B.~Moraes, F.B.~Abdalla, A.~Cuceu, M.~McLeod, L.~Whiteway et~al.,
  \emph{Cosmological measurements from angular power spectra analysis of boss
  dr12 tomography}, {\emph{Mon.\ Not.\ R.\ Astron.\ Soc.} {\bfseries 485}
  (2019) 326}.

\bibitem{wolz2016intensity}
L.~Wolz, C.~Tonini, C.~Blake and J.~Wyithe, \emph{Intensity mapping
  cross-correlations: connecting the largest scales to galaxy evolution},
  {\emph{Mon.\ Not.\ R.\ Astron.\ Soc.} {\bfseries 458} (2016) 3399}.

\bibitem{2011/sobreira}
F.~Sobreira, F.~de~Simoni, R.~Rosenfeld, L.~da~Costa, M.~Maia and M.~Makler,
  \emph{Cosmological forecasts from photometric measurements of the angular
  correlation function}, {\emph{Phys. Rev. D} {\bfseries 84} (2011) 103001}.

\bibitem{mcleod2017joint}
M.~McLeod, S.T.~Balan and F.B.~Abdalla, \emph{A joint analysis for cosmology
  and photometric redshift calibration using cross-correlations}, {\emph{Mon.\
  Not.\ R.\ Astron.\ Soc.} {\bfseries 466} (2017) 3558}.

\bibitem{2011/lesgourgues}
J.~Lesgourgues, \emph{The cosmic linear anisotropy solving system (class) i:
  Overview}, {\emph{arXiv preprint arXiv:1104.2932} (2011) }.

\bibitem{2011/blas}
D.~Blas, J.~Lesgourgues and T.~Tram, \emph{The cosmic linear anisotropy solving
  system {(CLASS). Part II}: approximation schemes}, {\emph{J. Cosmol.
  Astropart. Phys.} {\bfseries 2011} (2011) 034}.

\bibitem{alonso2019unified}
D.~Alonso, J.~Sanchez, A.~Slosar and L.D.E.S.~Collaboration, \emph{A unified
  pseudo-{Cl} framework}, {\emph{Mon.\ Not.\ R.\ Astron.\ Soc.} {\bfseries 484}
  (2019) 4127}.

\bibitem{hivon2002master}
E.~Hivon, K.M.~G{\'o}rski, C.B.~Netterfield, B.P.~Crill, S.~Prunet and
  F.~Hansen, \emph{Master of the cosmic microwave background anisotropy power
  spectrum: a fast method for statistical analysis of large and complex cosmic
  microwave background data sets}, {\emph{Astrophys.\ J.} {\bfseries 567}
  (2002) 2}.

\bibitem{mericia2023testing}
E.J.~Mericia, L.C.~Santos, C.A.~Wuensche, V.~Liccardo, C.P.~Novaes,
  J.~Delabrouille et~al., \emph{Testing synchrotron models and frequency
  resolution in bingo 21 cm simulated maps using gnilc}, {\emph{Astron.\
  Astrophys.} {\bfseries 671} (2023) A58}.

\bibitem{novaes2022bao}
C.P.~Novaes, J.~Zhang, E.J.~de~Mericia, F.B.~Abdalla, V.~Liccardo,
  C.A.~Wuensche et~al., \emph{The {BINGO project-VIII}. recovering the {BAO}
  signal in hi intensity mapping simulations}, {\emph{Astronomy \&
  Astrophysics} {\bfseries 666} (2022) A83}.

\bibitem{2023/Novaes_ML}
C.P.~Novaes, E.J.~de~Mericia, F.B.~Abdalla, C.A.~Wuensche, L.~Santos,
  J.~Delabrouille et~al., \emph{Cosmological constraints from low redshift 21
  cm intensity mapping with machine learning}, {\emph{Monthly Notices of the
  Royal Astronomical Society} {\bfseries 528} (2024) 2078}.

\bibitem{zhang2022transitioning}
Z.~Zhang, C.~Chang, P.~Larsen, L.F.~Secco, J.~Zuntz and L.D.E.S.~Collaboration,
  \emph{Transitioning from stage-iii to stage-iv: cosmology from galaxy$\times$
  cmb lensing and shear$\times$ cmb lensing}, {\emph{Mon.\ Not.\ R.\ Astron.\
  Soc.} {\bfseries 514} (2022) 2181}.

\bibitem{LSST:2018}
{The LSST Dark Energy Science Collaboration}, R.~Mandelbaum, T.~Eifler,
  R.~Hložek, T.~Collett, E.~Gawiser et~al., \emph{The lsst dark energy science
  collaboration (desc) science requirements document},  2021.

\bibitem{LSSTHandBook2011}
R.A.~Shaw and M.A.~Strauss, \emph{{LSST} data challenge handbook - version 1},
  2011.

\bibitem{2005/gorski}
K.M.~Gorski, E.~Hivon, A.J.~Banday, B.D.~Wandelt, F.K.~Hansen, M.~Reinecke
  et~al., \emph{Healpix: A framework for high-resolution discretization and
  fast analysis of data distributed on the sphere}, {\emph{Astrophys.\ J.}
  {\bfseries 622} (2005) 759}.

\bibitem{xavier2016flask}
H.S.~{Xavier}, F.B.~{Abdalla} and B.~{Joachimi}, \emph{{Improving lognormal
  models for cosmological fields}},
  \href{https://doi.org/10.1093/mnras/stw874}{\emph{Mon.\ Not.\ R.\ Astron.\
  Soc.} {\bfseries 459} (2016) 3693}
  [\href{https://arxiv.org/abs/1602.08503}{{\ttfamily 1602.08503}}].

\bibitem{zhang2022bingoVI}
J.~Zhang, P.~Motta, C.P.~Novaes, F.B.~Abdalla, A.A.~Costa, B.~Wang et~al.,
  \emph{The {BINGO project-VI}. {H{\sc i}} halo occupation distribution and
  mock building}, {\emph{Astron.\ Astrophys.} {\bfseries 664} (2022) A19}.

\bibitem{dunkley2009five}
J.~Dunkley, E.~Komatsu, M.~Nolta, D.~Spergel, D.~Larson, G.~Hinshaw et~al.,
  \emph{Five-year wilkinson microwave anisotropy probe* observations:
  likelihoods and parameters from the wmap data}, {\emph{Astrophys.\ J.\ Supp.}
  {\bfseries 180} (2009) 306}.

\bibitem{liccardo2022bingo}
V.~Liccardo, E.J.~de~Mericia, C.A.~Wuensche, E.~Abdalla, F.B.~Abdalla,
  L.~Barosi et~al., \emph{The {BINGO project-IV}. simulations for mission
  performance assessment and preliminary component separation steps},
  {\emph{Astron.\ Astrophys.} {\bfseries 664} (2022) A17}.

\bibitem{fornazier2021bingoV}
K.S.~Fornazier, F.B.~Abdalla, M.~Remazeilles, J.~Vieira, A.~Marins, E.~Abdalla
  et~al., \emph{The bingo project-v. further steps in component separation and
  bispectrum analysis}, {\emph{Astron.\ Astrophys.} {\bfseries 664} (2022)
  A18}.

\bibitem{delabrouille2013pre}
J.~Delabrouille, M.~Betoule, J.-B.~Melin, M.-A.~Miville-Desch{\^e}nes,
  J.~Gonzalez-Nuevo, M.~Le~Jeune et~al., \emph{The pre-launch planck sky model:
  a model of sky emission at submillimetre to centimetre wavelengths},
  {\emph{Astron.\ Astrophys.} {\bfseries 553} (2013) A96}.

\bibitem{2015/remazeilles}
M.~Remazeilles, C.~Dickinson, A.~Banday, M.-A.~Bigot-Sazy and T.~Ghosh,
  \emph{An improved source-subtracted and destriped 408-mhz all-sky map},
  {\emph{Mon.\ Not.\ R.\ Astron.\ Soc.} {\bfseries 451} (2015) 4311}.

\bibitem{2008/miville-deschenes}
M.-A.~Miville-Desch{\^e}nes, N.~Ysard, A.~Lavabre, N.~Ponthieu,
  J.-F.~Macias-Perez, J.~Aumont et~al., \emph{Separation of anomalous and
  synchrotron emissions using wmap polarization data}, {\emph{Astron.\
  Astrophys.} {\bfseries 490} (2008) 1093}.

\bibitem{2003/dickinson}
C.~Dickinson, R.~Davies and R.~Davis, \emph{Towards a free--free template for
  cmb foregrounds}, {\emph{Mon.\ Not.\ R.\ Astron.\ Soc.} {\bfseries 341}
  (2003) 369}.

\bibitem{planck2016planck}
P.~Collaboration et~al., \emph{Planck 2015 results: Xxvi. the second planck
  catalogue of compact sources}, {\emph{Astron.\ Astrophys.} {\bfseries 594}
  (2016) A26}.

\bibitem{tinker2010large}
J.L.~Tinker, B.E.~Robertson, A.V.~Kravtsov, A.~Klypin, M.S.~Warren, G.~Yepes
  et~al., \emph{The large-scale bias of dark matter halos: numerical
  calibration and model tests}, {\emph{Astrophys.\ J.} {\bfseries 724} (2010)
  878}.

\bibitem{abdalla2022bingo3}
F.B.~Abdalla, A.~Marins, P.~Motta, E.~Abdalla, R.M.~Ribeiro, C.A.~Wuensche
  et~al., \emph{The {BINGO Project-III}. optical design and optimization of the
  focal plane}, {\emph{Astron.\ Astrophys.} {\bfseries 664} (2022) A16}.

\bibitem{Remazeilles:2011}
M.~{Remazeilles}, J.~{Delabrouille} and J.-F.~{Cardoso}, \emph{{CMB and SZ
  effect separation with constrained Internal Linear Combinations}},
  \href{https://doi.org/10.1111/j.1365-2966.2010.17624.x}{\emph{Mon.\ Not.\ R.\
  Astron.\ Soc.} {\bfseries 410} (2011) 2481}
  [\href{https://arxiv.org/abs/1006.5599}{{\ttfamily 1006.5599}}].

\bibitem{2021/cunnington}
S.~Cunnington, M.O.~Irfan, I.P.~Carucci, A.~Pourtsidou and J.~Bobin,
  \emph{21-cm foregrounds and polarization leakage: cleaning and mitigation
  strategies}, {\emph{Mon.\ Not.\ R.\ Astron.\ Soc.} {\bfseries 504} (2021)
  208}.

\bibitem{wolz2017determining}
L.~Wolz, C.~Blake and J.~Wyithe, \emph{Determining the h i content of galaxies
  via intensity mapping cross-correlations}, {\emph{Mon.\ Not.\ R.\ Astron.\
  Soc.} {\bfseries 470} (2017) 3220}.

\bibitem{wolz2019intensity}
L.~Wolz, S.~Murray, C.~Blake and J.~Wyithe, \emph{Intensity mapping
  cross-correlations ii: Hi halo models including shot noise}, {\emph{Mon.\
  Not.\ R.\ Astron.\ Soc.} {\bfseries 484} (2019) 1007}.

\bibitem{Jiang:2023}
Y.-E.~Jiang, Y.~Gong, M.~Zhang, Q.~Xiong, X.~Zhou, F.~Deng et~al.,
  \emph{Cross-correlation forecast of csst spectroscopic galaxy and meerkat
  neutral hydrogen intensity mapping surveys},
  \href{https://doi.org/10.1088/1674-4527/accdc0}{\emph{Research in Astronomy
  and Astrophysics} {\bfseries 23} (2023) 075003}.

\bibitem{mazumder2025hi}
A.~Mazumder, L.~Wolz, Z.~Chen, S.~Paul, M.~Santos, M.~Jarvis et~al., \emph{{HI
  intensity mapping with the MIGHTEE} survey: First results of the {HI} power
  spectrum}, {\emph{arXiv preprint arXiv:2501.17564} (2025) }.

\bibitem{zheng2025cosmology}
J.~Zheng, P.~Tiwari, G.-B.~Zhao, D.J.~Schwarz, D.~Bacon, S.~Camera et~al.,
  \emph{Cosmology from lofar two-metre sky survey data release 2:
  Cross-correlations with luminous red galaxies from eboss}, {\emph{arXiv
  preprint arXiv:2504.20722} (2025) }.

\bibitem{virtanen2020scipy}
P.~Virtanen, R.~Gommers, T.E.~Oliphant and et~al., \emph{Scipy 1.0: Fundamental
  algorithms for scientific computing in python},
  \href{https://doi.org/10.1038/s41592-019-0686-2}{\emph{Nat.\ Methods}
  {\bfseries 17} (2020) 261}.

\bibitem{hartlap2007why}
J.~Hartlap, P.~Simon and P.~Schneider, \emph{Why your model parameter
  confidences might be too optimistic. unbiased estimation of the inverse
  covariance matrix}, {\emph{Astron.\ Astrophys.} {\bfseries 464} (2007) 399}.

\bibitem{whelan2018astropy}
A.M.~Price-Whelan, B.~Sip{\H{o}}cz, H.~G{\"u}nther, P.~Lim, S.~Crawford,
  S.~Conseil et~al., \emph{The astropy project: building an open-science
  project and status of the v2. 0 core package}, {\emph{Astron.\ J.} {\bfseries
  156} (2018) 123}.

\bibitem{zonca2019healpy}
A.~Zonca, L.~Singer, D.~Lenz, M.~Reinecke, C.~Rosset, E.~Hivon et~al.,
  \emph{healpy: equal area pixelization and spherical harmonics transforms for
  data on the sphere in python},
  \href{https://doi.org/10.21105/joss.01298}{\emph{J. Open Source Softw.}
  {\bfseries 4} (2019) 1298}.

\bibitem{2011/numpy}
S.~Van Der~Walt, S.C.~Colbert and G.~Varoquaux, \emph{The numpy array: a
  structure for efficient numerical computation}, {\emph{Computing in science
  \& engineering} {\bfseries 13} (2011) 22}.

\bibitem{2007/matplotlib}
J.D.~Hunter, \emph{Matplotlib: A 2d graphics environment}, {\emph{Comput. Sci.
  Eng.} {\bfseries 9} (2007) 90}.

\end{thebibliography}\endgroup
\bibliographystyle{JHEP}

%\begin{thebibliography}{99}

%\bibitem{a}
%Author, \emph{Title}, \emph{J. Abbrev.} {\bf vol} (year) pg.

%\bibitem{b}
%Author, \emph{Title},
%arxiv:1234.5678.

%\bibitem{c}
%Author, \emph{Title},
%Publisher (year).

% Please avoid comments such as "For a review'', "For some examples",
% "and references therein" or move them in the text. In general,
% please leave only references in the bibliography and move all
% accessory text in footnotes.

% Also, please have only one work for each \bibitem.

%\end{thebibliography}
\end{document}